\documentclass[aip,pof,preprint]{revtex4-2}
\usepackage{graphicx}
\usepackage{array}[=2016-10-06]
\usepackage{tikz}
\usepackage{subcaption}
\usepackage{tabularx}
\usepackage{booktabs}
\usepackage[english]{babel}
\usepackage{bm}
\usepackage{dsfont}
\usepackage{amsfonts,amsmath,amssymb}
\usepackage{booktabs}
\usepackage{verbatim}
\usepackage{cancel}

\usepackage[justification=raggedright,singlelinecheck=false]{caption}

\usetikzlibrary{calc}

\newcommand{\td}[2]{\frac{\mathrm{d} #1}{\mathrm{d} #2}}
\newcommand{\de}{\text{d}}
\usepackage{xr-hyper}
\usepackage[colorlinks=true,linkcolor=teal,citecolor=blue]{hyperref}

\begin{document}

\title{Flow-induced bending response rheometer to measure viscoelastic bending of  microrods}

\author{Barrett T Smith}
\affiliation{Chemical Engineering, Northeastern University}

\author{Michał Czerepaniak}
\affiliation{Faculty of Physics, University of Warsaw, Poland}

\author{Maciej Lisicki}
\affiliation{Faculty of Physics, University of Warsaw, Poland}

\author{Sara M Hashmi}
\email[]{s.hashmi@northeastern.edu}
\affiliation{Chemical Engineering, Northeastern University}
\affiliation{Mechanical \& Industrial Engineering, Northeastern University}
\affiliation{Chemistry \& Chemical Biology, Northeastern University}

\date{\today}

\begin{abstract}

Soft, microscale hydrogel fibers and rods play important roles in tissue engineering, flexible electronics, soft robotics, drug delivery, sensors, and other applications.  Their viscoelastic mechanical properties, while critical for their function, can be challenging to characterize.  We present a flow-induced bending response (FIBR) rheometer that quantifies the bending modulus and viscoelastic properties of small, hydrated fibers and rods using flow through a glass capillary.  The fiber is positioned across the capillary entrance, and pressure-driven, controlled inflow of water exerts a quantifiable force on the sample.  Fiber deflection is determined by video microscopy obtained simultaneously with measurements of flow rate.  We develop an analytical model to resolve the hydrodynamic forces applied to the rod, and use Euler-Bernoulli beam theory to determine its material properties.  Using a constant volume flow rate of water enables measurement of steady rod deflection, and thus the bending modulus.  Application of viscous forces to the rod in a stepwise, cyclic or oscillatory manner enables measurement of time-dependent responses, creep recovery, viscoelastic moduli, and other properties.  We demonstrate the versatility of this technique on natural and synthetic materials spanning diameters from 5 to 300  microns and elastic moduli ranging from 100 Pa to $>$100 MPa.  Because the technique uses water to exert forces on the fiber, it works particularly well for hydrated materials, such as hydrogels and biological fibers, providing a versatile platform to characterize microscale mechanical properties of elongated structures.
\end{abstract}


\maketitle

\section{Introduction}

Soft, microscale, hydrogel fibers and small rods ($<200$ $\mu$m in diameter) play important roles in tissue engineering, flexible electronics, soft robotics, drug delivery, sensors, and many other applications \cite{du_hydrogel_2023, hertle_naturally_2024, li_recent_2023, xi_soft_2017}.
Their viscoelastic mechanical properties are critical for their function, but characterizing these properties can be challenging.  Bulk and macroscale testing, such as shear rheology or traditional dynamic mechanical analysis techniques, is often unfeasible for small, soft, fibrous materials as they are often too small to load into macroscale equipment. Further, particularly soft samples may be too soft to be effectively clamped into place, as the clamping force may damage the material.  Additionally, biological samples and hydrogels have different properties in wet and dry environments. These limitations necessitate alternative techniques that can probe microscale materials, especially in their hydrated state.

Atomic force microscopy (AFM) with Force Spectroscopy Mapping is a well-established and widely used nanoindentation method for determining the mechanical properties of small, soft materials \cite{heinz1999spatially, garcia2020nanomechanical}. The ability to perform and accurately interpret AFM experiments depends on the probe shape, size, and sample interactions.
The probe size and shape are critical to the success of the measurement, and each requires separate mathematical models to correctly interpret the results.
Probe geometry determines the amount and nature of the contact between the probe and sample surface. Therefore, different probe shapes experience different adhesion forces and varying degrees of interference from the underlying substrate \cite{chyasnavichyus_recent_2015, gisbert_accurate_2021, tsui_substrate_1999, vichare_cellular_2013}.
Larger probes match the bulk properties more precisely but are more prone to substrate interference \cite{asgari_revealing_2022}. Smaller probes show more variability over the material surface, particularly for porous materials such as hydrogels \cite{asgari_revealing_2022, abrego_multiscale_2021}. As a result, bulk material properties do not always match localized AFM measurements \cite{asgari_revealing_2022, flores-merino_nanoscopic_2010, solares_nanoscale_2016}.
In addition to probe-related considerations, the aqueous environment required by most hydrogels and biological samples adds further complications.
Surface tension complicates measurements, particularly for soft materials, where the forces involved are of a similar magnitude to the surface tension \cite{ding_determination_2017, pham_elasticity_2017}. Friction and adhesion issues are also more pronounced in water \cite{he_comprehensive_2024, megone_impact_2018}. This can make it difficult to find the material surface and add to the complexity of the analysis.
While AFM can provide reasonable mechanical measurements of soft materials when experiments are carefully designed and controlled, its complexity is nontrivial and motivates the development of alternative, simpler, and more accessible approaches.

Beyond AFM, many techniques have been developed for characterizing the material properties of small, soft materials \cite{salipante_microfluidic_2023}. Optical tweezers can probe very small hydrated samples to determine their viscoelastic mechanical properties. However, the generated forces may be insufficient for stiffer samples, and samples must be transparent \cite{robertson-anderson_optical_2024, catala-castro_exploring_2022}.
Another approach involves fixing one or both ends of a fiber in place and using gravity or fluid flow to induce bending while measuring the resulting deformation \cite{amir_bending_2014, duprat2015microfluidic, chakrabarti_instabilities_2021}. This approach has been demonstrated on several materials, each requiring a unique experimental design.  However, the required sample configuration is not feasible for all sample types. In one instance, a fluidic device is custom-designed with wells cut into the sides of a primary channel to allow bacteria to diffuse into them.  Growth of bacteria, with one end fixed in the well and the other extending into the main channel, allows the transverse fluid flow to bend the extended portion \cite{amir_bending_2014}.  In another instance, photopolymerized polymer filaments are formed \textit{in situ} in a microfluidic channel so that each filament extends beyond the width of the main flow channel and each end is held in place by the geometry of the chamber.  The filaments are thereby automatically oriented perpendicular to the flow, which bends them.  However, filaments must be fabricated \textit{in situ} for this method to work \cite{duprat2015microfluidic}.  In a third approach, long extruded polymer fibers are hung over the edge of a platform and gravity bends them.  However, extrusion generally forms fibers no smaller than a few hundred microns in diameter.  Further, this method has been demonstrated in air: hydrated samples, especially smaller ones, may experience hydrodynamic forces that are difficult to avoid \cite{chakrabarti_instabilities_2021}.
Capillary aspiration techniques provide an alternative by using glass capillaries that form a seal on the material surface, with an applied vacuum causing measurable deformation \cite{aoki_pipette_1997, guo_micromechanics_2011, mwyss_capillary_2010}. In capillary micromechanics, materials suspended in pressure-driven flows are immobilized at the tapered end of a capillary. Deformation as a function of applied stress provides mechanical property measurements \cite{mwyss_capillary_2010}.  Both capillary aspiration and capillary micromechanics techniques are frequently used on spherical samples.  Achieving a tight seal may be difficult for especially porous or irregular surfaces, resulting in errors in measurement or difficulty in generalizing the mechanical properties of a small piece of the surface to an entire material.
A related approach eliminates the need for a seal by relying instead on fluid flow past the sample to exert the deforming force, although this technique has only been demonstrated to measure the elastic modulus of diatom colonies \cite{young_quantifying_2012}. Further validation and theoretical development is needed to establish and expand this method.

To address the shortcomings of available bulk and microscale characterization techniques, we propose a flow-induced bending response (FIBR) rheometer to determine the bending modulus and viscoelastic properties of small hydrated fibers and rods using flow through a glass capillary. The technique positions the fiber perpendicular to the capillary.  The flow of water into the capillary exerts quantifiable forces on the sample, and microscopy measures the fiber deflection to determine the material moduli. The method is simple and uses common readily available equipment.
It works well for fibers, both natural and synthetic, with diameters between 1 and 500 $\mu$m. We measure materials with elastic moduli ranging over nearly 7 orders of magnitude in stiffness, from 100 Pa to $>$100 MPa. Because the technique uses water to exert force on the fiber, it works particularly well for hydrated materials, such as hydrogels and biological fibers.
We develop a theoretical analytical framework that accounts for hydrodynamic interactions between the fiber and the fluid flow to calculate the force exerted on the fiber.  Then we use Euler-Bernoulli bending theory to determine the bending modulus as a function of the applied force and measured fiber deflection.  Here we provide a detailed validation of the FIBR rheometer and show its ability to go beyond elastic measurements to measure time-dependent and viscoelastic material properties as well as investigating failure.

\section{Materials and Methods}

\subsection{Materials}

Sterilized downy duck feathers are taken from a couch pillow (Pacific Coast Feather Cushion, North Carolina USA). The barbules are stripped from the barbs via tweezers. Polyester fibers are obtained from the mouthpiece plug of a VWR Disposable Serological Pipet and used without further modification.

Sodium alginate, magnesium chloride, and sodium chloride are obtained from Sigma-Aldrich. Calcium chloride is obtained from VWR. All chemicals are used as received. Solutions of sodium alginate, calcium chloride, magnesium chloride, and sodium chloride are prepared in DI water (Milli-Q Advantage A10). We obtain fluoresceinamine from Thermo Scientific to label the alginate used in the alginate rod fabrication described below.  Briefly, 255.2mg sodium alginate, 18.5mg fluoresceinamine (Thermo Scientific), and 35.9mg N-(3-Dimethylaminopropyl)-N-ethylcarbodiimide hydrochloride (Sigma) are dissolved in phosphate buffered solution. The reaction proceeds for 24 hours, followed by dialyzation in DI water, and lyophilization into a dry powder before use, according to the following literature procedures \cite{scheja2017glucose, smith_diffusion-driven_2024}. Concentration details are provided in the following sections.

Glass capillaries with internal diameters of 200, 600, 1000, or 2000 $\mu$m are purchased from VitroCom and World Precision Instruments (WPI) and are connected to pump tubing via a capillary holder (WPI MPH3) or, for capillaries that do not fit in the holder, are attached directly to the pump tubing using epoxy.

\subsection{Alginate Fiber Manufacture}

Alginate fibers are generated by extrusion through a blunt needle into a calcium bath \cite{zhou_continuous_2019, cuadros_mechanical_2012}. Sodium alginate is dissolved in water at a concentration of 2\% by weight. A syringe pump pushes 0.5 mL of the alginate solution through a 34 gauge blunt needle at 30 mL/hr into a bath of calcium chloride (2\% in water). The alginate fibers form instantly, although gelation continues for at least 24 hr before any measurements to allow for the complete diffusion of calcium through the fiber. The fibers produced by this method have diameters between 125 and 250 $\mu$m.

To significantly change the mechanical properties of the alginate fibers, we use magnesium as an alternate crosslinker or include sodium chloride in the ion bath, as both are known to soften alginate gels \cite{topuz_magnesium_2012, leroux_compressive_1999}.
However, we find that direct extrusion into a magnesium bath does not produce fibers; only calcium chloride at specific weight percents (1-4\%) allows for the formation of fibers.
This is likely because magnesium binds much more weakly with alginate than does calcium, and that magnesium reacts with alginate much more slowly than calcium \cite{topuz_magnesium_2012}.
Therefore, the fibers are removed from the calcium bath with tweezers immediately after extrusion and placed into a bath with alternate salts: either a 2\% solution of magnesium chloride or a solution with 2\% calcium chloride and 1\% sodium chloride. While some calcium carries over with the newly formed alginate fibers, we transfer only approximately 0.1 g of fiber into a 10 mL bath, so that the calcium concentration is reduced by at least a factor of 100.
Fibers are allowed to equilibrate in the new ion solution for at least 72 hr before measurement. Fibers placed in the magnesium bath swell slightly for a final diameter of 200–300 $\mu$m. Fibers cross-linked in a calcium bath are referred to as Ca-alginate fibers, while those transferred to the magnesium bath are Mg-alginate fibers.

\subsection{Alginate Rod Manufacture}

Alginate rods are generated in a microfluidic device \cite{smith_situ_2024, smith2024diffusion}. The device consists of a Y-shaped intersection. All channels are 40 $\mu$m wide and 25 $\mu$m tall. A pressure controller (Fluigent LU-FEZ) with an attached flow-rate meter is used to control flow rates. A solution of fluorescently-tagged sodium alginate (0.1 mg/mL) flows at 1 $\mu$L/min in one inlet and meets a solution of calcium chloride (100 mM) also flowing at 1 $\mu$L/min. Sodium chloride is present in both streams at 2.5 mM. Where the two inlet streams meet at the Y-intersection, gel deposits on the channel wall. The gel grows to fill the cross-section of the channel for up to 1 mm downstream of the junction. The deposited gel occludes the channel, requiring higher driving pressures to maintain the set flow rate. At a certain threshold, shear stresses from the fluid are sufficient to remove the gel from the channel wall and flush it out of the device. The eluted gel keeps the form of the channel, resulting in a rod 500-1000 $\mu$m long with a $25\times40$ $\mu$m rectangular cross-section. Alginate rods are kept in the elutant until measured.

\subsection{Rheology}

Bulk alginate gels are made according to an established method \cite{agulhon_influence_2014}. Sodium alginate solution (2\% in water) is added to a 3D printed circular mold (50 mm in diameter). Filter paper (Whatman no. 1) impregnated with 2\% calcium chloride solution is placed atop the alginate solution and secured in place with another 3D printed part. Several milliliters of a solution of the desired divalent cation is placed on top of the filter and allowed to diffuse into the alginate solution. After at least one hour, the gel can be removed from the mold, the filter stripped from the gel, and the gel placed in a larger volume of the desired ion solution, either 2\% calcium chloride or 4\% magnesium chloride. Gelation is allowed to continue for at least one day.  After the alginate has completely gelled, the sample is trimmed to the size of the rheometer geometry.

Prepared gels are added to a 40 mm plate-plate geometry for small angle oscillatory shear (SAOS) measurements. The plate is lowered onto the sample until the normal force reaches 0.1 N, setting a gap of approximately 3 mm. The measurements consist of a frequency sweep followed by an amplitude sweep.
The amplitude sweep is measured at an angular frequency $\omega=5$ rad/s.  All samples exhibit a linear regime with constant modulus at strains below 0.1\%.
The frequency sweep is measured at an oscillatory strain of 0.1\%, which is within the linear regime found from the amplitude sweep for all samples.

\subsection{ FIBR Rheometer}

Glass capillaries with internal diameters of $200-2000$ $\mu$m are secured into the tubing of a pressure controller with an associated flow rate meter (Fluigent LineUp Push-Pull).  The pump can operate between $-0.9$ bar and 1 bar.
A custom designed 3D-printed well (4 mL capacity) is affixed to a glass slide using epoxy. The capillary is positioned using a micromanipulator (WPI) over the objective of an inverted microscope (Leica DMIL). The entire setup is mounted on an air table (Thorlabs) to minimize vibrations.
A schematic of the experimental setup is provided in Figure \ref{fig:experimental setup}.

The glass slide well is filled with water before adding a sample. The sample position is manipulated using the microscope stage and/or tweezers until the rod is positioned transversely to the capillary opening. A small negative pressure draws water into the capillary, holding the rod in position before the flow tests begin. This corresponds to 50-500 $\mu$L/min. The required flow rate depends on fiber and capillary size.
To maintain a constant volume of water in the slide well during all measurements, a separate pump supplies water to the well at a flow rate equal to that being removed through the capillary.

We position the sample across the opening of the capillary along the diameter of the capillary, by applying a gentle vacuum, so that each end of the sample contacts the capillary wall.  To verify the sample is centered relative to the center of the capillary opening, we adjust the focal plane to the center of the capillary.  Using either $10\times$ or $20\times$ objectives on the microscope enables a depth of field either $\pm2.3$ or $\pm 0.23$  $\mu$m, respectively. These depth of field options allow us to position the sample in the center of the capillary to within a few \% of the fiber diameter, even for the smallest samples.  If the edges of the capillary and the edges of the fiber are simultaneously in focus, the fiber is centered within the capillary. If, however, the capillary edges are in focus but the fiber is not, then the fiber is off-center and must be repositioned. We confirm that the fiber is settled in place by moving the capillary and observing the fiber move with it.

Once the rod is positioned, flow tests are conducted. We control applied pressure and record flow rate $Q$ through the Fluigent software for the duration of the experiment. We perform three different types of flow tests: elastic measurements, creep measurements, and dynamic mechanical analysis (DMA). For each, the increment size and maximum applied vacuum  depend on the sample's effective stiffness.
The appropriate applied pressure for any sample is not known a priori, but is determined via experimentation.
The achievable flow rates depends on the size of the capillary. For the small 200 $\mu$m capillary, flow rates vary between $0-1000$ $\mu$L/min, while for the large 2000 $\mu$m capillary, achievable flow rates exceed 12000 $\mu$L/min.

For measurements of elastic materials, vacuum is applied in regular increments. At least several seconds elapse between each increment to allow the sample to reach steady-state  deflection. For creep measurements, a large step change in flow rate is applied and held for more than 60 s while deflection is measured. Afterwards, flow rate is dropped to a baseline of sufficient magnitude to prevent the fiber from falling off the capillary.
For DMA measurements, flow rate is varied in a sinusoidal pattern with a period of 10 s.

To maintain the sample's position throughout all tests described above, we maintain a flow into the capillary at all times, where even the minimum flow applied is sufficient to hold the sample on the capillary. Slip is sometimes, but rarely, observed under the microscope, in some of the larger deflection measurements.  However if we notice much slippage, we exclude the data from the analysis.

\subsection{Microscopy and Video Analysis}

We record fiber deflection $\delta$ using a bright field inverted microscope (Leica DMIL) at either $10\times$ or $20\times$ magnification, at a spatial resolution of 0.77 and 1.53 pixels/$\mu$m respectively. The microfluidic alginate rods are imaged using fluorescent microscopy, as they contain a fluorescein labeled alginate. The timestamp from each video frame is matched to the corresponding flow rate data to determine the deflection $\delta$ and flow rate $Q$ for each frame.

Each video frame is analyzed to determine the position of the rod or fiber edge using the steps indicated briefly in Figure \ref{fig:VidAnalysis} and in more detail in Figure \ref{fig:SIVidAnalysis}. First, images are cropped to focus on the rod edge positioned in the center of the capillary where we expect the most deflection. When samples are fluorescent, fiber edges are easily identified.  For non-fluorescent samples,  grayscale images are binarized. For samples that are transparent or for which binarization alone does not reveal the rod edge, we use image filters provided in the scikit-image package in Python. Of the options within the package, the Sato or Sobel filters provide the cleanest results  to illuminate the edges before binarization. The Sato filter, in particular, is designed to measure tubes, ridges, or wrinkles, and is most successful across our samples \cite{van_der_walt_scikit-image_2014}. 
Having successfully isolated the rod edges, we find all contours in the image and sort them by $x$-position to determine the right and left edges of the rod. The leading, inside edge of the sample is used, which is in contact with the capillary for the duration of the measurement. The inside and outside edge of the fibers deflect together for nearly all materials. Only the softest, thickest fibers show some discrepancy between the inside and outside edge, which we suspect is compression in the fiber. In the largest Mg-alginate fibers, the inside edge deflects slightly more than the outside edge: these differences are <50\%. Finally, a 10th order polynomial, sufficient to capture any irregularities in the rod edge, is fitted to the contour for use when comparing across frames. After identifying a baseline frame, polynomial fits from each frame are compared to the baseline at each $x$-position to find the greatest distance between the edges. This distance is the deflection of the rod $\delta$. Figure \ref{fig:SIVidAnalysis} provides the entire image analysis procedure demonstrated on a non-fluorescent sample for which binarization alone is not sufficient to identify the edges and the Sato filter is used.

\begin{figure}[htbp]
    \begin{subfigure}[b]{0.49\textwidth}
        \resizebox{\linewidth}{!}{
            \def\skipstandalone{}
            \ifdefined\skipstandalone
\else
    \documentclass[aip,pof,graphicx,JoR,preprint]{revtex4-1}
    \usepackage{tikz}
    \usetikzlibrary{calc}
    \begin{document}
\fi
\definecolor{capillary}{RGB}{120,160,180}
\definecolor{capillaryend}{RGB}{90,130,150}
\definecolor{capillarygradient}{RGB}{150,180,200}
\definecolor{pumpcolor}{RGB}{80,100,120}
\definecolor{tubecolor}{RGB}{100,120,140}
\definecolor{rodcolor}{RGB}{140,100,80}
\definecolor{wellcolor}{RGB}{180,210,230}
\definecolor{wellborder}{RGB}{80,110,140}
\definecolor{objectivebody}{RGB}{100,110,120}
\definecolor{objectivelens}{RGB}{150,190,210}
\begin{tikzpicture}
    \def\pumpx{1.5}
    \def\pumpy{2}
    \def\pumpround{0.25cm}
    \def\tubethick{1.5mm}
    \def\caplen{3}
    \def\capdiam{0.5}
    \def\rodwidth{2}
    \def\rodheight{0.2}
    \def\rodround{0.1cm}
    \def\arrowthick{1.5mm}
    \def\arrowlen{1}
    \def\capangle{-30}
    \def\wellwidth{4.5}
    \def\wellheight{2}
    \def\wellborderwidth{5pt}
    \def\objwidth{1.2}
    \def\objheight{1.5}
    \def\objtop{0.6}
    \def\captaper{0.5}
    
    \fill[pumpcolor, rounded corners=\pumpround] (0,0) rectangle (\pumpx,\pumpy) coordinate (pumpright) at (\pumpx,\pumpy/2);
    
    \draw[tubecolor, line width=\tubethick] ($(pumpright)+(-\pumpx/2, \pumpy/2)$) .. controls ++(0.5,1) and ++(0,0.5).. ++(\pumpx/2+0.5,0.8) coordinate (tubeend);
    
    \begin{scope}[rotate around={\capangle:(tubeend)}, shift={(tubeend)}]
        \fill[capillary] (0,0) coordinate (capleft) circle (\captaper*\capdiam);
        
        \fill[capillary] ($(capleft)+(0,\captaper*\capdiam)$) -- 
            ++(\caplen,{(1-\captaper)*\capdiam}) -- 
            ++(0,-2*\capdiam) -- 
            ++(-\caplen,{(1-\captaper)*\capdiam}) -- 
            cycle;
        
        \fill[capillaryend] (capleft) ++(\caplen,0) coordinate (capright) circle (\capdiam);
        \fill[capillarygradient] (capright) circle (0.75*\capdiam);
        
        \draw[-to, black, line width = \arrowthick] ($(capright)+(\arrowlen+0.6,0)$) -- +(-\arrowlen,0);
    \end{scope}
    
    \fill[rodcolor, rounded corners = \rodround] ($(capright)+(-\rodwidth/2,\rodheight/2)$) coordinate (rodcenter) rectangle ++(\rodwidth,-\rodheight);
    
    \fill[wellcolor, opacity=0.5] ($(capright)+(-\wellwidth/2*1.1,-\wellheight/2*1.25)$) coordinate (wellleft) rectangle ++(\wellwidth,\wellheight) coordinate (wellright) coordinate (wellbottom) at (wellleft);
    \draw[wellborder, line width=\wellborderwidth, line join=round] ($(wellleft)+(0,\wellheight)$) -- (wellleft) -- ++(\wellwidth,0) -- ++(0,\wellheight);

    \coordinate (objbottom) at ($(wellbottom)+(\wellwidth/2*1.1,-\wellheight)$);
    
    \fill[objectivebody] ($(objbottom)+(-\objwidth/2,0)$) rectangle ($(objbottom)+(\objwidth/2,2*\objheight/3)$);
    
    \fill[objectivebody] ($(objbottom)+(-\objwidth/2,2*\objheight/3-0.01)$) -- 
        ($(objbottom)+(\objwidth/2,2*\objheight/3-0.01)$) -- 
        ($(objbottom)+(\objtop/2,\objheight)$) -- 
        ($(objbottom)+(-\objtop/2,\objheight)$) -- cycle;

    
    \fill[objectivelens] ($(objbottom)+(0,\objheight)$) ellipse (\objtop/2 and 0.1);
    \draw[objectivebody, line width=1pt] ($(objbottom)+(0,\objheight)$) ellipse (\objtop/2 and 0.1);

    \path (wellright) ++(0.5,0) coordinate (resupplyX);
    \path (pumpright) ++(0,-\pumpy/2) coordinate (resupplyYBottom);
    \fill[pumpcolor, rounded corners=\pumpround] (resupplyX |- resupplyYBottom) rectangle ++(\pumpx,\pumpy);
    
    \coordinate (resupplyTop) at ($(resupplyX |- resupplyYBottom)+(\pumpx/2,\pumpy)$);
    
    \draw[-to, tubecolor, line width=\tubethick] ($(resupplyTop)$) .. controls ++(-1,1.5) and ++(0,1) .. ($(wellright)+(-0.8,-0.4)$);

    
    \node[anchor=north, font=\small, align=center] at ($(\pumpx/2,0)$) {Vacuum\\[-8pt]Flow\\[-8pt]Controller};
    
    \node[anchor=south, font=\small, rotate={\capangle+atan((1-\captaper)*\capdiam/\caplen)}] at ($(capleft)+(\caplen/2.5,-0.4)$) {Glass Capillary};
    
    \node[anchor=center, font=\small] at ($(rodcenter)+(-0.5,-0.1)$) {Fiber};
    
    \node[anchor=south, font=\small, rotate=\capangle] at ($(capright)+(\arrowlen/2+0.7,-0.6)$) {Flow};
    
    \node[anchor=west, font=\small] at ($(wellleft)+(0, -0.3)$) {Well};
    
    \node[anchor=north, font=\small, align=center] at ($(objbottom)$) {Microscope\\[-8pt]Objective};


    \node[anchor=north, font=\small, align=center] at ($(resupplyTop)+(0,-\pumpy)$) { Water\\[-8pt]Resupply\\[-8pt]Pump};
    
\end{tikzpicture}
\ifdefined\skipstandalone
\else
    \end{document}
\fi
        }
        \vspace{-10pt}
        \caption{}
        \label{fig:setup}
    \end{subfigure}
    \hfill
    \begin{subfigure}[b]{0.49\textwidth}
        \includegraphics[width=\linewidth]{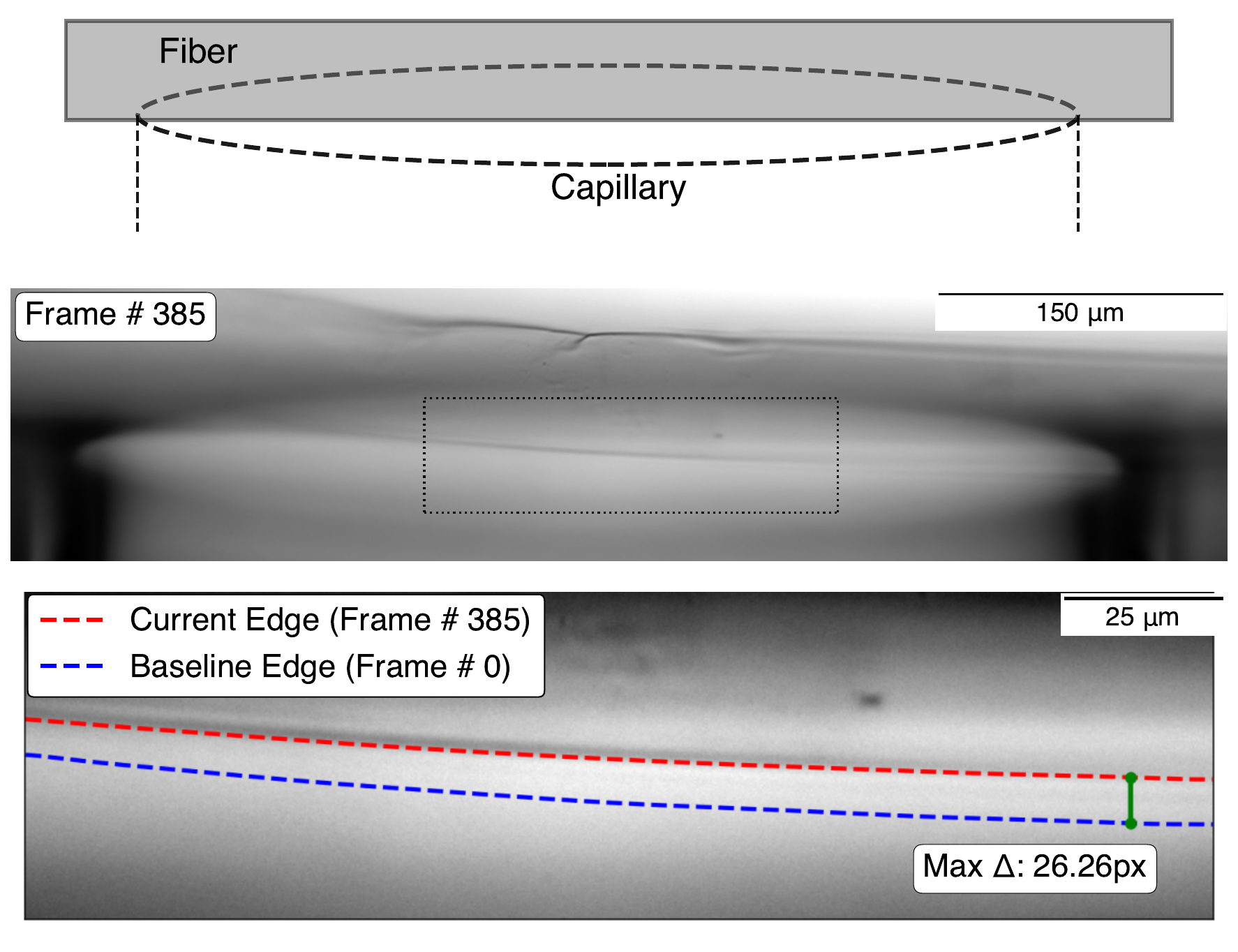}
        \caption{}
        \label{fig:VidAnalysis}
    \end{subfigure}
    \caption{(a) Schematic of the experimental setup.
    (b) Example microscopy and image analysis.  The top shows a schematic representation of the fiber positioned atop the capillary in a microscopy image.  The middle shows a raw microscope image in which the entire diameter of the capillary is visible and an alginate fiber sample is in place across the capillary opening.  The dotted box represents the region of interest cropped for analysis.  The bottom image shows a comparison of the edge found in this frame 385, in red, to the baseline, zero-deflection frame 0, in blue.  The green line segment shows the greatest difference between the two contours.}
    \label{fig:experimental setup}
\end{figure}

\section{Theory}

The fluid drag forces that act on a fiber placed in an external flow induce a deformation, the shape of which results from the balance between elastic and hydrodynamic stresses. Given the microscale sizes, the flow around the fiber is governed by the Stokes equations. To resolve the hydrodynamic forces,  we construct a bead model and use the mobility matrix formalism. The details of the calculation are included in the SI. We approximate hydrodynamic interactions between the subunits of the fiber with the Rotne-Prager-Yamakawa tensor\cite{rotne1969variational,yamakawa1970transport,zuk2014rotne}. This approach enables us to relate the local drag acting on the filament to the imposed velocity distribution at the entrance to the capillary tube. These forces are then used to model the deflection using a continuous Euler-Bernoulli beam equation~\cite{bauchau2009structural}.

We assume that the deformation or bending deflection $\delta$ of the fiber is sufficiently small such that the change in shape does not affect the form of the mobility matrix. In practice, this assumption means that $\delta/2R$ is small, where $R$ is the radius of the capillary.  Table~\ref{tab:ReTable} in the SI provides measurements of the maximum deflection $\delta_{max}$ measured for each material, along with the smallest $R$ used to measure that material. The largest measured deflections are up to 95~$\mu$m for fibers tested on the 2000~$\mu$m diameter capillary, 13~$\mu$m for fibers tested with the 600~$\mu$m capillary, and 30~$\mu$m for fibers tested with the 200~$\mu$m capillary. In all materials measured, $\delta_{max}/2R \lesssim 5$\%, except for the alginate rods measured with the 200~$\mu$m capillary, in which $\delta_{max}/2R\sim15\%$. After presenting the small-deformation theory, we will revisit this assumption in describing the experimental results.

The capillary entrance flow is not fully developed and evolves rapidly from a plug-like profile at the entrance-region inlet toward the parabolic Hagen--Poiseuille profile inside the tube \cite{white1999fluid,gross2021mathematical}. We therefore use a simple interpolating ansatz for the axisymmetric flow profile at the location of the fiber,
\begin{equation}
    U_\text{ext}=U_\text{max}\left[1-\left(\frac{r}{R}\right)^4\right],
    \label{eq:quartic_flow}
\end{equation}
where $r$ is the distance from the axis and $R$ is the capillary radius, and we neglect the disturbance of the imposed flow by the fiber itself.
The exponent 4 is not intended to represent the exact asymptotic entrance profile; rather, it provides a tractable profile that is flatter than fully developed Poiseuille flow, while still satisfying the no-slip condition at the wall. To assess the sensitivity of the model to this assumption, we repeat the elastic calculation using a steeper profile,
$U_\text{ext}=U_\text{max}[1-(r/R)^6]$.
For a general exponent $n$, the imposed flow rate satisfies
$Q=\pi R^2 U_\text{max} n/(n+2)$. Changing from $n=4$ to $n=6$ modifies the proportionality between $Q$ and $U_\text{max}$ by approximately 11\%. However, the resulting $F_\text{max}$ vs. $U_\text{max}$ relationship changes by less than 4\%. Therefore, the choice of exponent mainly affects numerical prefactors and does not alter the main conclusions of the model. We expect the neglected disturbance of the entrance flow by the fiber itself to introduce a larger uncertainty than the choice between exponents 4 and 6.
The velocity field is related to the imposed volumetric flow rate inside the capillary
\begin{equation} \label{eq:flow}
    Q = \int U_\text{ext} \,\de S = \int_0^{2\pi}\int_0^R U_\text{max}\left(1 - \frac{r^4}{R^4} \right) \,r \de r \de\phi = \frac{2}{3}\pi R^2 U_\text{max}.
\end{equation}

The theoretical approach assumes low Reynolds numbers.  For each experiment we define a fiber-scale Reynolds number
\begin{equation}
\mathrm{Re}_f=\frac{\rho U_{\max} d}{\eta},
\end{equation}
with \(U_{\max}=3Q/(2\pi R^2)\) from Eq.~\eqref{eq:flow}. Each experiment requires a range of flow rates and thus covers a range of \(\mathrm{Re}_f\) values. The maximum values for the materials used in Table~\ref{tab:Bending Results} are reported in Table~\ref{tab:ReTable} in the SI. For the largest fibers and highest flow rates only, \(\mathrm{Re}_f\) is no longer asymptotically small, and reaches up to \(\mathrm{Re}_f=41\).  The present hydrodynamic model should be interpreted as a leading-order approximation for \(\mathrm{Re}_f>O(1)\).

\begin{figure}
    \centering
    \includegraphics[width=1\linewidth]{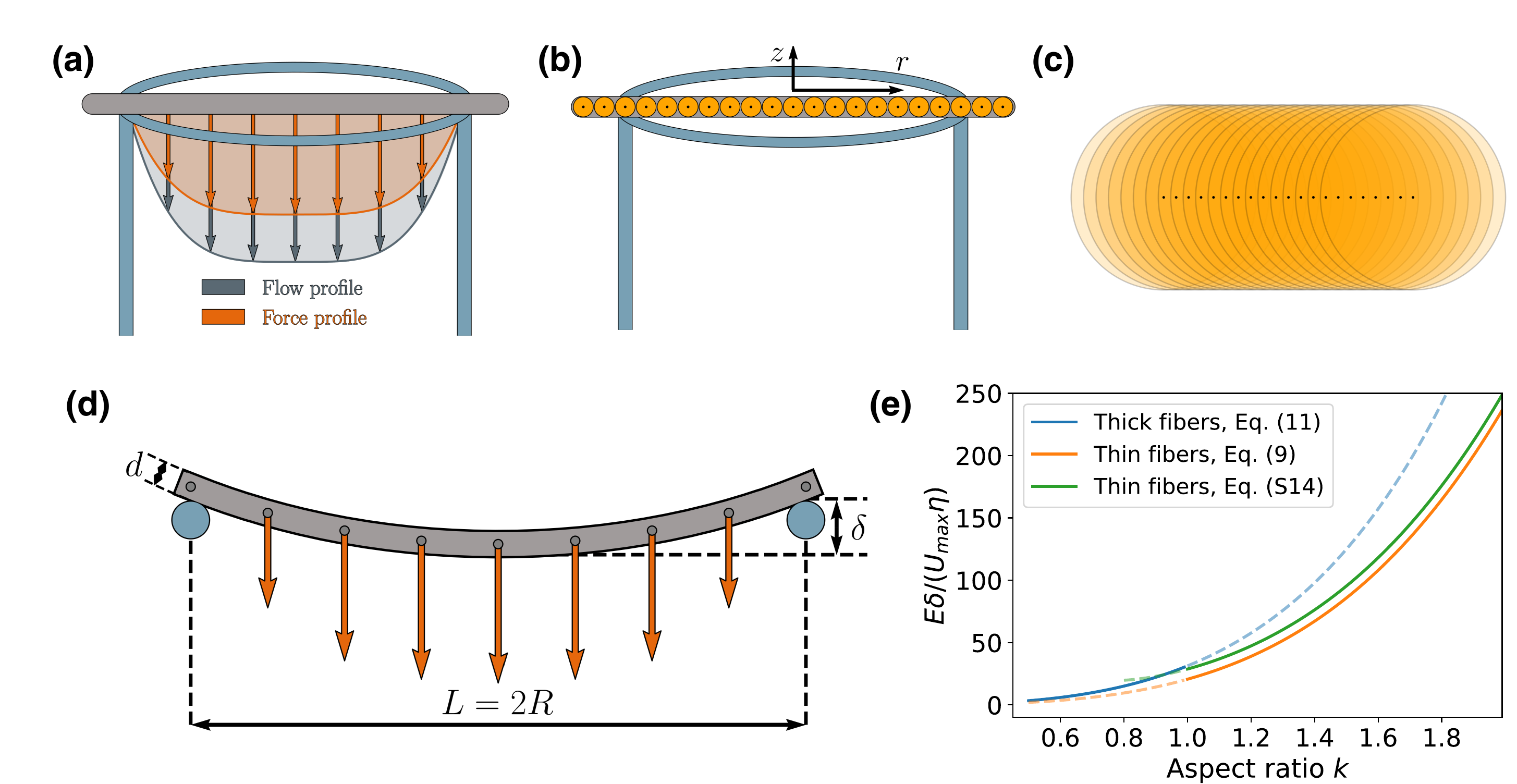}
    \caption{(a) Schematic illustration of a fiber (grey) placed inside a capillary tube, showing the flow velocity profile, Eq.~\eqref{eq:quartic_flow}, and the force profile acting on the fiber, Eq.~\eqref{eq:force_quartic}. (b) A continuous ``thin'' fiber is discretized as a collection of mutually connected beads, each pair of neighboring beads being in point contact. This approach allows the computation of hydrodynamic forces acting on the fiber using
    Eq.~\eqref{eq:RPY_ij_thin}.
    (c) A continuous ``thick'' fiber is discretized as a set of overlapping, connected beads. This approach allows the computation of hydrodynamic forces acting on the fiber using
    Eq.~\eqref{eq:RPY_ij_thick}.
    (d) Schematic illustration of a deformed fiber: the capillary tube has radius $R$, the fiber has thickness $d$, and the deflection is denoted by $\delta$. The distribution of hydrodynamic forces acting on the fiber is also sketched.
    (e) Plot of the dimensionless Young's modulus $E \delta/U_{max}\eta$,  as a function of the ratio of the tube radius to the fiber thickness, $k=R/d$, for 'thin' fibers (Eq.~\eqref{eq:E_thin_cylinder}) in orange and 'thick' fibers (Eq.~\eqref{eq:E_thick_cylinder}) in blue. The dashed lines indicate the continuation of the relationship, extrapolated into regimes where it is no longer valid (e.g., the 'thick' regime extended into the 'thin' regime). The modulus plotted in green represents the exact expression (Eq.~\eqref{eq:E_exact}) for thin fibers, relaxing the assumption of large aspect ratio $k$.}
    \label{fig:theory_fig}
\end{figure}

To introduce elastic stresses, we discretize the fiber into beads of radius $a$ (and diameter $d$) each, with the distance $l_0$ between their centers. Next, we assume that the associated profile of the drag force density $f$ (force per unit length) has a shape that is qualitatively similar to that of the flow,
\begin{equation}
    f = \frac{F_\text{max}}{l_0}\left[1-\left(\frac{r}{R}\right)^4\right],
    \label{eq:force_quartic}
\end{equation}
where $l_0$ is the distance between the centers of the neighboring beads. Both force and velocity profiles are sketched in Figure \ref{fig:theory_fig}a. We expect this assumption to hold in the central part of the fiber, close to the axis of the capillary. This assumption can be justified using arguments analogous to classical resistive force theory, where the relationship between the velocity and force density remains a purely local quantity~\cite{gray1955propulsion}.

Once the magnitude of the force is known, the maximal deflection of a deformed fiber that is accessible in experiments can be found from solving the Euler-Bernoulli elastic beam problem in the limit of small deflections. In the present geometry, the relevant small-deflection parameter is the ratio of the transverse displacement to the support span, $\delta/(2R)$, together with the requirement that the beam slope remains small everywhere. Under these conditions, geometric nonlinearities remain weak and linear Euler-Bernoulli theory provides an adequate description. Parameterizing the shape of the fiber by its arclength $s$, the vertical deflection $z(s)$ satisfies
\begin{equation}
    EI\td{^4z}{s^4} = - \frac{F_\text{max}}{l_0}\left(1-\frac{s^4}{R^4}\right ),
    \label{eq:euler_bernouli_beam}
\end{equation}
with $E$ being the Young's modulus, and $I$ being the second moment of cross-sectional area of the fiber. We solve this equation with the boundary conditions of the fiber being simply supported at the walls of the capillary, which means that its vertical position is fixed there, and the ends are torque-free, so that
\begin{equation}
    z\left(\pm R\right)  = 0 = z''(\pm R).
    \label{eq:euler_bernouli_bc}
\end{equation}

The choice of boundary conditions is guided by several observations and supported by previous studies of the bending of polyethylene glycol diacrylate (PEGDA) filaments in microfluidic channels \cite{duprat2015microfluidic}.  In the previous study, PEGDA filaments are formed in situ in a fluidic channel by photopolymerization through a mask, so that the filaments extend beyond the width of the main flow channel and are oriented perpendicular to the flow. A constant load is assumed along the transfer direction due to hydrodynamic pressure forces in the channel, and the filaments positioned across the channel are assumed to be simply supported at each end \cite{duprat2015microfluidic}.  In our experiments, the fibers remain fixed at the capillary edge. The measurement protocol ensures that the rods are securely immobilized prior to initiating the flow assay, consistent with a simply supported boundary condition at the contact points. At the same time, we observe that the fiber ends are able to rotate. Figure \ref{fig:SIRotatingEnds} shows two fields of view of alginate fiber ends in which the entire capillary opening is captured in the microscopy, at two different flow rates.   Accordingly, the analytical model treats the fiber as immobilized at the two contact points with the capillary while allowing for rotational freedom.  However, the precise nature of the contact can influence the inferred modulus, as more constrained, clamped-like conditions would increase the effective bending resistance. We note that when choosing such boundary conditions we neglect the influence of the remaining part of the fiber, outside of the capillary, on the overall shape of the fiber. However, we postulate that these parts would not dramatically influence the shape of the fiber in the central part of the capillary. We test Ca-alginate fibers of different total lengths, 12~mm and 1.2~mm, using a capillary with inner diameter 600 $\mu$m.  We obtain similar moduli for the two samples: 3.1 and 2.8~kPa, respectively.  This result suggests that the external fiber segments do not alter the measured bending response over this range, even when the fiber is 20 times longer than the capillary inner diameter.

Integration of Eq.~\eqref{eq:euler_bernouli_beam}  with the boundary conditions~\eqref{eq:euler_bernouli_bc} leads to the following relationship between the maximal deflection $\delta$ and the Young's modulus
\begin{equation}
    EI\delta \approx \frac{f_{max}R^4}{5}
    \label{eq:euler_bernoulli_EI},
\end{equation}
with $f_{max} = F_{max}/l_0$ being the maximal force per unit length of the fiber.

To relate the maximal force per unit length
to the imposed velocity amplitude $U_\text{max}$, we must consider hydrodynamic interactions within the fiber. To do so, we analyze the hydrodynamic drag on the mid-point of the fiber under a known distribution of hydrodynamic forces that results from the fiber being placed in an external flow. The details of this calculation rely on the mobility matrix formalism and are described in the SI. Here, we present the final result.

We distinguish two cases, depending on the fiber thickness $d$ in comparison to the radius of the capillary $R$. We define `thick' fibers as ones with a diameter greater than the radius of the capillary, while `thin' corresponds to an increasing aspect ratio $k=R/d$. The distinction between the modeling approaches for these two fiber regimes is illustrated in Figures \ref{fig:theory_fig}b and \ref{fig:theory_fig}c. For thin fibers, we find that
\begin{equation}
    f_\text{max} \approx U_\text{max} \frac{4\pi\eta}{\log{k} + 1.86}.
    \label{eq:force_thin}
\end{equation}
It is worth noting that the result is very close to that obtained using resistive force theory \cite{gray1955propulsion,johnson1979flagellar,dhont1996introduction}; it differs only by an additive term in the denominator, which becomes negligible for sufficiently large length-to-thickness ratios. This result is approximate, as it is based on the assumption of a sufficiently large aspect ratio. Nevertheless, the deviation from the result obtained without this assumption
(Eq.~\eqref{eq:force_exact})
is small.
For thick fibers, the resulting expression is
\begin{equation}
     f_\text{max} \approx U_\text{max}\frac{3\pi\eta}{\displaystyle \frac{R}{d}\left(\frac{8}{5}-\frac{3}{8}\frac{R}{d} \right)}
     \label{eq:force_thick}
\end{equation}
Because the hydrodynamic force scales with velocity, the force profile along the length of the fiber has the same shape as the flow profile, as illustrated in Figure \ref{fig:theory_fig}d.

Combining the results above, we find the expressions for the Young's modulus of the fibers. For a thin fiber with a cylindrical cross-section, with $I=\pi a^4/4$, and aspect ratio $k$, we have

\begin{equation}
    E = \frac{256}{5}\frac{U_\text{max}\eta}{\delta}\frac{k^4}{\log k+1.86},
    \label{eq:E_thin_cylinder}
\end{equation}

For non-circular cross-sections, the bead model remains the same, but the moment of interia changes.  For instance, for a rod with a square cross-section, $d\times d$ and $I=d^4/12$, we find
\begin{equation}
    E = \frac{48\pi}{5}\frac{U_\text{max}\eta}{\delta}\frac{k^4}{\log k+1.86}
    \label{eq:E_thin_square}
\end{equation}
For a thick cylindrical fiber, the logarithmic term is replaced and the expression reads
\begin{equation}
    E = \frac{ \eta U_\text{max}}{\delta}\frac{24 k^3}{\displaystyle 1-\frac{15}{64}k},
    \label{eq:E_thick_cylinder}
\end{equation}
while for a thick square cross-section,
\begin{equation}
    E = \frac{ \eta U_\text{max}}{\delta}\frac{9\pi k^3}{\displaystyle 2
    \left(1-\frac{15}{64}k\right)}.
    \label{eq:E_thick_square}
\end{equation}

The final expressions above can be used to directly predict the Young's modulus given the measured deflection of the fiber $\delta$ and its thickness $d$, as well as the size of the capillary used for the measurement. Figure \ref{fig:theory_fig}e plots $E(k)$ for thin (Eq.~\eqref{eq:E_thin_cylinder}) and thick (Eq.~\eqref{eq:E_thick_cylinder}) fibers, along with the exact solution for thin fibers, relaxing the assumption of large $k$ (Eq.~\eqref{eq:E_exact}).  At the limiting aspect ratio of $k=1$, the two approaches give similar results, thus together they provide a complete description for the analysis of small-deflection experimental data.

In all the cases considered, the Young's modulus is proportional to the ratio of maximal flow velocity to maximal deflection, and can be written as
\begin{equation}
    E=B(R,d)\frac{\eta Q}{\delta},
\end{equation}
where $B(R,d)$ is a geometric parameter, and the measured quantity $Q$ is derived from $U_\text{max}$ using Eq.~\eqref{eq:flow}. These expressions serve as the basis for the analysis and interpretation of experimental data throughout this work.

\section{Experimental Results}
\subsection{Behavior of elastic materials in the capillary bending assay}
\begin{figure}[htbp]
    \centering
    \includegraphics[width=1\linewidth]{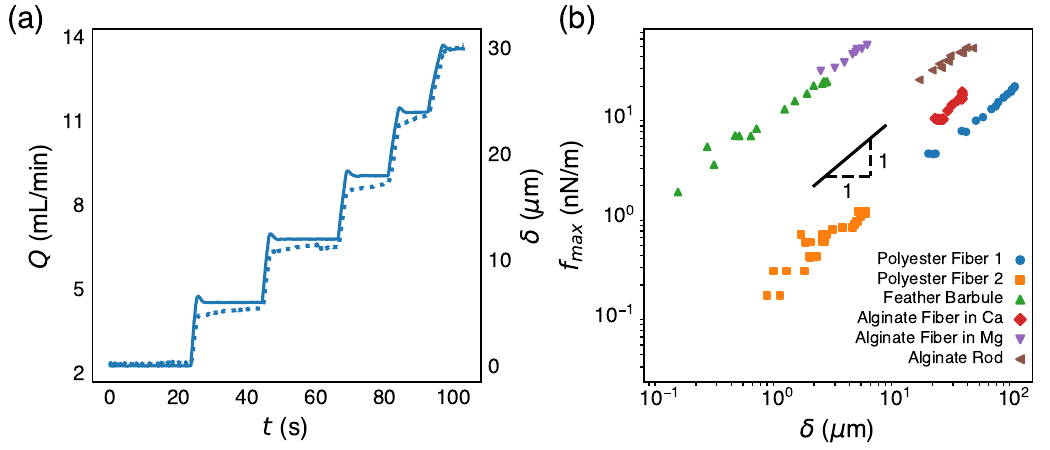}
    \caption{(a) Elastic behavior of a polyester fiber. The solid line shows the flow rate of water pulled into the capillary, while the dotted line shows the deflection of the fiber. (b) Elastic behavior in a range of materials, where the polyester fiber from (a) is in blue.}
    \label{fig:elastic}
\end{figure}

Figure \ref{fig:elastic}a shows the raw data for a polyester fiber bent with the FIBR rheometer technique on a 2000 $\mu$m capillary. The plot shows the volumetric flow rate $Q$ (left axis, solid blue line) and the fiber deflection $\delta$ (right axis, dotted orange line) as functions of time. The experiment starts at a flow rate of $\sim$2 mL/min which exerts a force sufficient to hold the fiber in place and prevent the fiber from falling from the center of the capillary. In the experiment, flow rate increases stepwise by 2 mL/min every $\sim$20 s.  The delay in changing flow rates ensures that the fiber reaches an equilibrium position at each new flow rate.  With each increase in flow rate, the fiber deflection increases a proportional amount, $\sim 6$ additional micrometers for each 2 mL/min increase in flow rate.

Figure \ref{fig:elastic}b shows measurements of deflection $\delta$ of several materials as a function of the force per unit length applied by the fluid flow.  Each data point represents a deflection similar to that seen in the plateaus in Figure \ref{fig:elastic}a.  We use Equation \ref{eq:force_thin} coupled with Equation \ref{eq:flow} to calculate
$f_\text{max}$ from measurements of $Q$. The plot is on log-log axes to demonstrate the wide range of possible exerted forces ($0.1-100$ nN) and measured deflections ($0.1-100$ $\mu$m).  The materials cover a range of synthetic and bio-materials.  Each material shows linear behavior, as indicated by a slope of 1 on the log-log plot. The polyester fiber in Figure \ref{fig:elastic}a is shown in blue in Figure \ref{fig:elastic}b.

The linear relationship between $f_\text{max}$ and $\delta$ suggests an elastic material response. Eq.~\eqref{eq:euler_bernoulli_EI} shows that $f_{max} \sim \delta EI/R^4$.  The slope of each measurement of $f_{max}(\delta)$ is proportional to the effective stiffness of the sample, $EI$.  Along with the sample dimensions, the slope can be used to determine the intrinsic elastic modulus of the material. The largest values of $\delta/(2R)$ occur for the alginate rods, shown as brown triangles in Fig.~\ref{fig:elastic}b. Table~\ref{tab:ReTable} in the SI shows that, for these samples, $\delta/(2R)\approx 0.15$ which is no longer asymptotically small.  However, the data in Fig.~\ref{fig:elastic}b remain close to linear over this range, indicating that the empirical proportionality between applied load and deflection is preserved within the experimental window studied here. Because we do not independently determine a material-specific onset of nonlinear response for each sample, we use this observed linearity as a practical criterion for the applicable linear regime.

We can see the effect of capillary size when comparing the two polyester fiber traces. Polyester fiber 1 (in blue) was measured on a capillary with $2R=2000$ $\mu$m, while polyester fiber 2 (orange) was measured on a capillary with $2R=600$ $\mu$m capillary. Both fibers are 28 $\mu$m in diameter. The achievable flow rates and forces are lower on the smaller capillary, and the resulting deflections are much smaller. Nevertheless, they share a line on the plot, indicating a similar effective stiffness. Therefore, since the two samples are the same size, with the same $I$, they share a similar modulus.

\begin{table}[htbp]
    \centering
    \small
    \setlength{\tabcolsep}{6pt}
    \renewcommand{\arraystretch}{1.3}
    \begin{tabularx}{\columnwidth}{
        >{\raggedright\arraybackslash}X
        c
        >{\raggedright\arraybackslash}X
        >{\raggedright\arraybackslash}X
        >{\raggedright\arraybackslash}X
    }
        \toprule[1.5pt]
        Material & $n$ & $EI$ (Pa m$^4$) & $E$ (Pa) & Modulus from other technique (Pa)\\
        \midrule[1pt]
        Downy feather barbule & 3 & $4.14\pm0.16\times10^{-15}$ & $1.1\pm0.04\times10^{8}$ & $E = 0.5$--$2.5\times10^9$, Ref.\cite{bonser_youngs_1995, bonser_influence_2001, reddy_structure_2007, fuller_structure_2015} \\
        Polyester Fiber & 4 & $5.9\pm1.55\times10^{-13}$ & $2.29\pm0.18\times10^{7}$ & $E = 1$--$10\times10^9$, Ref.\cite{plushchik_effects_2000, fraga_relationship_2003} \\
        Ca-alginate Fiber (2\%:2\%) & 4 & $3.58\pm2.38\times10^{-12}$ & $2.05\pm1.03\times10^{4}$ & $G'=5.0\times10^3$ (SAOS) \\
        Ca-alginate Fiber (2\%:2\%) + 1\%NaCl & 5 & $5.76\pm0.56\times10^{-12}$ & $1.47\pm0.46\times10^{4}$ & --- \\
        Mg-alginate Fiber (2\%:2\%) & 8 & $6.39\pm4.91\times10^{-14}$ & $4.82\pm3.05\times10^{2}$ & $G'=129$ (SAOS) \\
        Alginate Rods & 5 & $8.72\pm7.0\times10^{-16}$ & $4.55\pm3.65\times10^{4}$ & $E = 7\times10^3$ (AFM) \\
        \bottomrule[1.5pt]
    \end{tabularx}
    \caption{The effective stiffnesses and elastic moduli for many materials as tested via the FIBR rheometer. All values were obtained using the thin-fiber model, unless otherwise noted in the text. Also, elastic or shear moduli for similar materials via other techniques.}
    \label{tab:Bending Results}
\end{table}

Using Eq.~\eqref{eq:euler_bernoulli_EI} we convert the slopes of the data shown in Figure~\ref{fig:elastic}b to measurements of $E$. Table~\ref{tab:Bending Results} shows several biological and synthetic materials measured using this FIBR rheometer technique.
Unless noted otherwise, all values reported in Table~\ref{tab:Bending Results} are extracted using the thin-fiber expressions, Eqs.~\eqref{eq:E_thin_cylinder} and \eqref{eq:E_thin_square}. The polyester fiber and feather barbule lie well within the thin-fiber regime, while the Mg-alginate fibers are closest to the crossover to the thick-fiber regime at $k=R/d\sim 1$. The smallest $k$ in our dataset belongs to a Mg-alginate fiber of 253 $\mu$m diameter measured on a 600 $\mu$m capillary ($k=1.18$). Using the thin fiber approximation, we find a modulus of 531 Pa, while the thick fiber approximation yields 590 Pa, a difference of 11\%.
The effective stiffness of the materials range over 6 orders of magnitude from $10^{-18}$ to $10^{-12}$ Pa m$^4$. The moduli also vary by more than six orders of magnitude from $10^2$ to $10^8$ Pa. Across the analyzed range of flow rates, we observe experimentally a linear relationship between the force, calculated from the imposed flow rate $Q$, and the measured deflection $\delta$ that is consistent with a linear elastic response of the fibers. Any substantial slippage or nonlinear behavior of the material would perturb this scaling.

Down feathers are widely used in clothing and upholstery, and their mechanical properties are well-characterized \cite{bonser_youngs_1995, bonser_influence_2001, reddy_structure_2007, fuller_structure_2015}. Feathers contain a central spine called a rachis, from which branch regularly spaced fibers known as barbs with diameters of $\sim$20 $\mu$m. Branching from the barbs are much smaller fibers called barbules, typically $\sim$2 $\mu$m in diameter and $<$0.5 mm in length. Barbs are sufficiently large to be measured using macroscale techniques and have Young's moduli ranging from $0.5-2.5$ GPa. A 20 $\mu$m feather barb with an elastic modulus of 2 GPa will have an effective stiffness $EI>10^{-11}$, which is too stiff for this technique. Feather barbules are too small for macroscale characterization, but share a chemical composition with the barbs. The smaller diameter of the feather barbules significantly lowers the effective stiffness so that they can be measured on the capillary. As measured on a 200 $\mu$m capillary, the feather barbules have a bending modulus of  110 MPa. We are not aware of any previous measurements of the bending modulus of individual feather barbules. The measured modulus is lower than that of feather barbs. Beyond the difference in materials (barb versus barbule), one reason for this difference may be the hydration of the barbule. The elastic moduli of feather barbs is reduced by nearly half when wet \cite{bonser_influence_2001, fuller_structure_2015}.

Polyesters are synthetic polymer materials that are used in a wide range of applications. The elastic modulus varies significantly based on the crystallinity and fiber alignment within the sample, but is generally between 1 and 10 GPa \cite{shirataki_correlation_1997,plushchik_effects_2000, fraga_relationship_2003}. The polyester fiber modulus measured via the FIBR rheometer, either with a 600 $\mu$m or 2000 $\mu$m capillary as shown in Figure \ref{fig:elastic}b, is significantly lower than that expected at  22.9 MPa. There are several possible explanations for this observation. First, similar to the feather barbule, the elastic modulus is significantly lower when the fibers are submerged in water \cite{fraga_relationship_2003}. Second, polyester fibers may be hollow, changing the area moment of inertia $I$ and skewing the elastic modulus calculation \cite{karaca_influence_2007}. The difference in bending rigidity between a filled and hollow fiber may become apparent if the fiber lumen is comparable to the diameter of the fiber. Both the area moment of inertia and bending rigidity of a full cylinder are proportional to the fourth power of the radius.  For a cylindrical annulus, the area moment of inertia is proportional to the difference of the fourth powers of the inner and outer radii. Therefore the difference in bending rigidity of a hollow cylinder compared to a filled one is negligible, less than $\sim6$\%, as long as the lumen is smaller than $\sim50$\% of the outer diameter.  In our measurements of the polyester fiber, microscopy suggests that any internal `lumen' is small compared to the fiber diameter.  One measurement suggests a lumen between 8-12 $\mu$m within a 28 $\mu$m polyester fiber.  Using $I$ for a hollow fiber would reduce the bending rigidity by only 0.7 - 3\%. This justifies the use of the filled area moment of inertia for a small density of hollow sites along the filament.

Alginate is a biopolymer extracted from sea algae used in food, cosmetics, and pharmaceuticals. Multivalent cations crosslink the polymer chains to form a gel. Alginate gels in the bulk have been widely studied and vary in their mechanical properties based on the specific composition and molecular weight of alginate and the species and concentration of the cross-linker.
The modulus of bulk calcium alginate gels can vary from 1 to $>$100 kPa \cite{leroux_compressive_1999, malektaj_mechanical_2023, larsen_rheological_2015}. Magnesium alginate gels are much softer, with elastic moduli ranging from 0.07--1.8 kPa \cite{topuz_magnesium_2012}.  Beyond changing the cross-linking ion, the moduli of alginate gels can be lowered by the addition of monovalent cations such as sodium \cite{leroux_compressive_1999}.  Differences in manufacturing processes also impact the resulting gel, with factors like molecular fiber alignment and post-gelation drying impacting the elastic modulus \cite{zhang_creating_2014, su_generation_2009}. Because of the wide variation in material properties, we perform small-amplitude oscillatory shear (SAOS) tests on bulk gels formed with either calcium (Ca-alginate) or magnesium (Mg-alginate) as the cross-linkers. When measured via SAOS, the Ca-alginate and Mg-alginate gels have storage moduli $G'$ of 5040 Pa and 124 Pa, respectively (see Figure \ref{fig:SIRheology}). These are well within the range of moduli found in the literature.

Table \ref{tab:Bending Results} lists the bending results for the three different compositions of alginate fibers measured on 600, 1000, or 2000 $\mu$m capillaries. Calcium alginate fibers have a measured elastic modulus of  20.5 kPa. The modulus decreases by $\sim$30\% to  14.7 kPa when sodium chloride is added to the calcium ion bath. Using Mg as the cross-linking agent results in the weakest fibers, with an elastic modulus of only  482 Pa. Because fibers formed via bath extrusion are likely isotropic and homogeneous \cite{zhang_creating_2014}, we can relate the bending modulus of the fibers found by the FIBR rheometer to the elastic modulus of the bulk gels determined via SAOS.
To compare the results from the rheometer and capillary bending, we must consider that the Young's modulus and storage component of the shear modulus are not the same measurement.
Instead, for isotropic materials with a Poisson's ratio of 0.5 and bulk modulus $B\gg E$, the relationship between the elastic modulus $E$ and shear modulus $G'$ is $E=3G'$   \cite{roylance_engineering_2001, lakes_viscoelastic_2017}.  For both the Ca-alginate and the Mg-alginate fibers, the moduli measured by the FIBR rheometer are indeed $\sim 3\times$ the storage moduli values of the bulk alginate gels as determined by SAOS. Thus, we see excellent agreement between the two separate measurement techniques.

Finally, we measure alginate rods produced in a microfluidic device. These rods are unique in several respects. They form at concentrations of alginate much lower than is possible by the direct extrusion method used above (0.01 \% versus 2 \%). The rods are formed during microfluidic flow, while cross-linking in the extruded fibers occurs only after the alginate flows through the needle, so that their internal structure is isotropic. Additionally, the alginate rods are microporous, whereas the extruded fibers are uniform. Because of their small size, $25\times40$ $\mu$m in cross-section and less than 1mm in length, bulk scale characterization is not possible.
As shown in Table \ref{tab:Bending Results}, the alginate rods are substantially stiffer than the alginate fibers at  $E=45.5$ kPa as measured on a 200 $\mu$m capillary. Interestingly, the alginate rods are softer when measured via AFM at 7 kPa. This is likely due to flow-induced internal fiber alignment within the alginate rods, which causes anisotropy in the material.
In a different context, alginate fibers made by electrospinning causes fiber alignment, also inducing anisotropic structure, as confirmed by small angle X-ray scattering \cite{zhang_creating_2014}. The anisotropic fibers have higher elastic moduli than their isotropic counterparts.

\subsection{Non-elastic behavior}

For many materials, an elastic modulus is sufficient to characterize the mechanical response. However, for viscoelastic materials, the FIBR rheometer can also characterize the time-dependent viscous response, either via creep measurements or Dynamic Mechanical Analysis techniques. We also explore fiber failure.

\subsubsection{Creep}

\begin{figure}[htbp]
    \centering
    \includegraphics[width=1\linewidth]{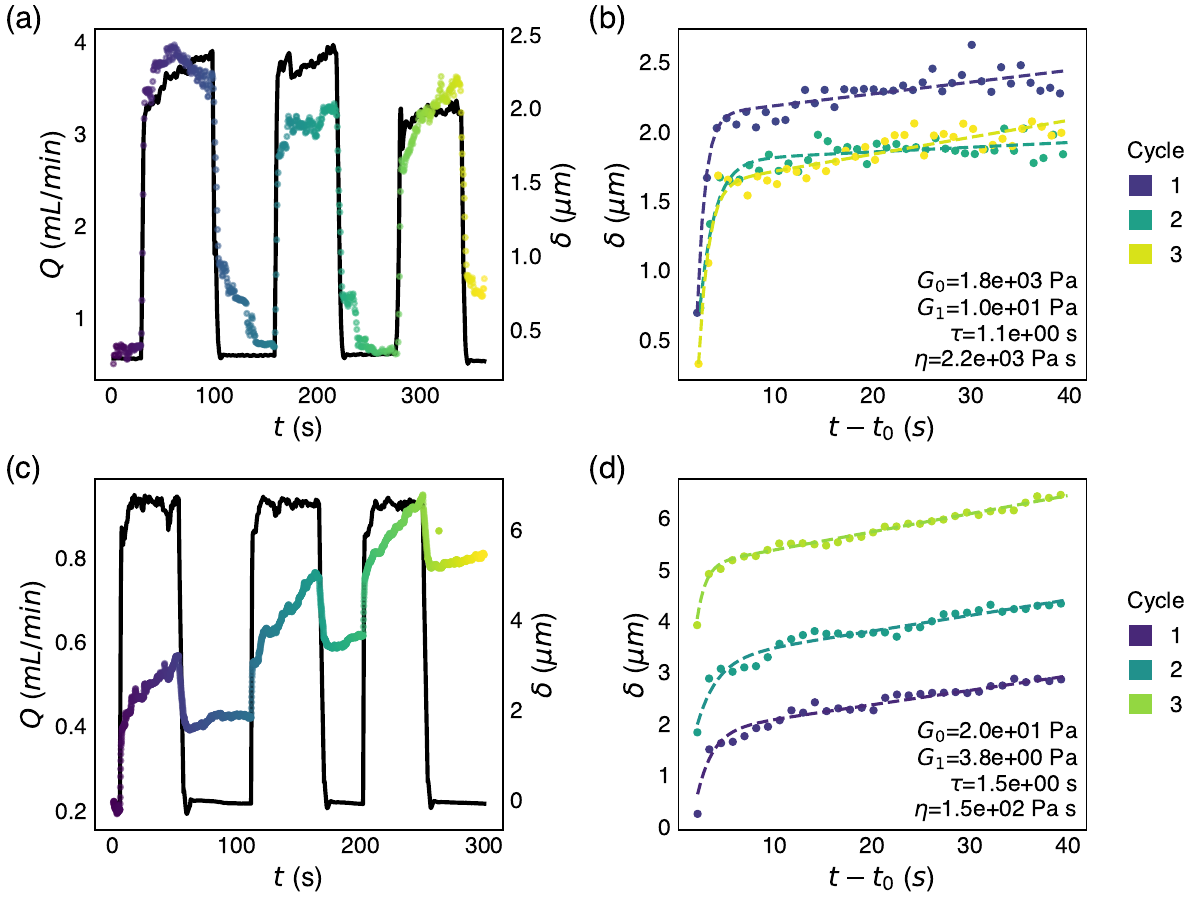}
    \caption{Cyclic loading reveals fiber creep. (a) The applied flow rate (black) and deflection (colored by time) of a Ca-alginate fiber during 3 square-wave loading cycles.  (b) Creep data for the 3 loading cycles from (a) fitted with the Burgers model (dashed lines). The average fit parameters are listed in the lower right.
    (c) The applied flow rate (black) and deflection (colored by time) of a Mg-alginate fiber during a 3 square-wave loading cycle. (d) Creep data for the 3 loading cycles from (c) fitted with the Burgers model (dashed lines). The average fit parameters are listed in the lower right.}
    \label{fig:Square Cycle Creep}
\end{figure}

Figure \ref{fig:Square Cycle Creep} shows the creep behavior of the different alginate fibers. Creep refers to the time-dependent movement of a material upon application of a steady force.  Figure  \ref{fig:Square Cycle Creep}a shows the raw flow rate (black) and deflection data (colored by time) for the Ca-alginate fiber creep measurements on a 1mm capillary. At $t=30, 160$ and 280 s, flow rate is increased and held steady for at least $60$ s to apply a constant force. Afterwards, the flow rate is dropped to a low value. At each application of force the fiber reaches an equilibrium deflection within several seconds. Upon removal of the force, the fiber recovers its original baseline deflection within 30 s.

Figure  \ref{fig:Square Cycle Creep}c shows a similar experiment for a Mg-alginate fiber measured on a 600 $\mu$m capillary. Because the Mg-alginate fiber is much less stiff, it deflects a similar amount on a smaller capillary and at lower flow rates than the Ca-alginate fiber measured on the 1000 $\mu$m capillary in Figure  \ref{fig:Square Cycle Creep}a. At each application of force, the Mg-alginate fiber in Figure  \ref{fig:Square Cycle Creep}c shows an initial 1.5 $\mu$m jump in deflection, followed by a linear increase in deflection, while the force is maintained. Upon releasing the force, the deflection drops an amount similar to the initial increase, but does not recover from the slower, linear deformation. As a result, there is a net increase in fiber deflection over the course of the creep cycles.

Figures  \ref{fig:Square Cycle Creep}b and \ref{fig:Square Cycle Creep}d show the creep data from Figures \ref{fig:Square Cycle Creep}a and \ref{fig:Square Cycle Creep}c, respectively, as well as the fit of the Burgers model of viscoelasticity\cite{malkin_rheology_2022, li_facile_2017}. The color of each curve in Figures \ref{fig:Square Cycle Creep}b and \ref{fig:Square Cycle Creep}d corresponds to the same colored cycle in Figures \ref{fig:Square Cycle Creep}a and \ref{fig:Square Cycle Creep}c. The Burgers model consists of a Kelvin-Voigt element, a damper and spring in parallel, in series with a Maxwell element, a damper and spring in series.  The damper elements model the viscous contribution to the mechanical response of the material, while the springs model the elastic contribution.  Purely elastic materials have a rapid response to creep test measurements, and reach their equilibrium shape quickly.  Materials with significant viscoelasticity take longer to recover. The viscoelastic compliance $J$ measures the ratio of strain to applied stress over time.  For  a Burgers material $J(t)$ behaves according to:

\begin{equation}
J(t) = \frac{\gamma(t)}{\sigma_0} = \frac{1}{G_0} + \frac{1}{G_1}\left(1-e^{-t/\tau}\right) + \frac{t}{\eta}
\end{equation}

\noindent where $G_0$ and $G_1$ are the shear moduli of the spring elements in the model, $\tau$ is the decay time, and $\eta$ is the material viscosity.  To calculate the viscoelastic compliance $J(t)$, we determine strain in the rod $\gamma(t)$ from the deflection: $ \gamma(t) = {3\delta d}/{2R^2}$, and stress in the rod $\sigma_0$ from the force: $\sigma_0 = 16 {FR}/{\pi d^3}$, using three-point flexural analysis based on the Euler Bernoulli beam bending theory. We note here that the analysis may be repeated with a distributed load model of Eq.~\eqref{eq:force_quartic} instead of a three-point force application. However, the resulting expressions for the stress and strain, $\gamma(t)=7 \delta d/ 6 R^2$ and $\sigma_0 = 28 FR/3\pi d^3$ differ only by the numerical prefactor of the same order of magnitude. With a distributed load, the predicted stress and strain are lower, as expected, by $\sim20$\% and $\sim40$\%, respectively.
For the purpose of fitting the model, each creep cycle is processed independently. The deflection at the start of each creep measurement is subtracted from all deflection values during that creep cycle so that the deflection starts at 0 for fitting. After fitting the model to this offset data, the initial deflection is added back to the fitted curves to generate the plots in Figures \ref{fig:Square Cycle Creep}b and \ref{fig:Square Cycle Creep}d .

Comparing the fit parameters found in Figures \ref{fig:Square Cycle Creep}b and \ref{fig:Square Cycle Creep}d between the Ca-alginate and Mg-alginate fibers shows dramatic differences. The instantaneous elastic response $G_0$ for the Ca-alginate fiber is approximately 100 times larger than $G_0$ for the Mg-alginate fiber, similar to the difference seen in the elastic moduli found in Table~\ref{tab:Bending Results}. This makes sense as the Ca-alginate fiber shows a similar magnitude deflection despite experiencing much faster flow, and consequently higher forces, on a larger capillary. The materials have similar Kelvin-Voigt elements ($G_1$ and $\tau$). The last dashpot element ($\eta$) shows that the material viscosity for the Ca-alginate fiber is an order of magnitude higher than for the Mg-alginate fiber, as is seen in the slow creep of the Mg-alginate fiber which is largely absent from the Ca-alginate fiber. The Kelvin-Voigt model on its own works reasonably well to describe the behavior of the Ca-alginate fiber, as seen in Figure \ref{fig:SI_KV_Creep}.  However, due to the slow creep in the Mg-alginate fiber, the Burgers model provides a better fit than Kelvin-Voigt (Figure \ref{fig:SI_KV_Creep}).

\subsubsection{Dynamic Mechanical Analysis}
\begin{figure}[htbp]
    \centering
    \includegraphics[width=1\linewidth]{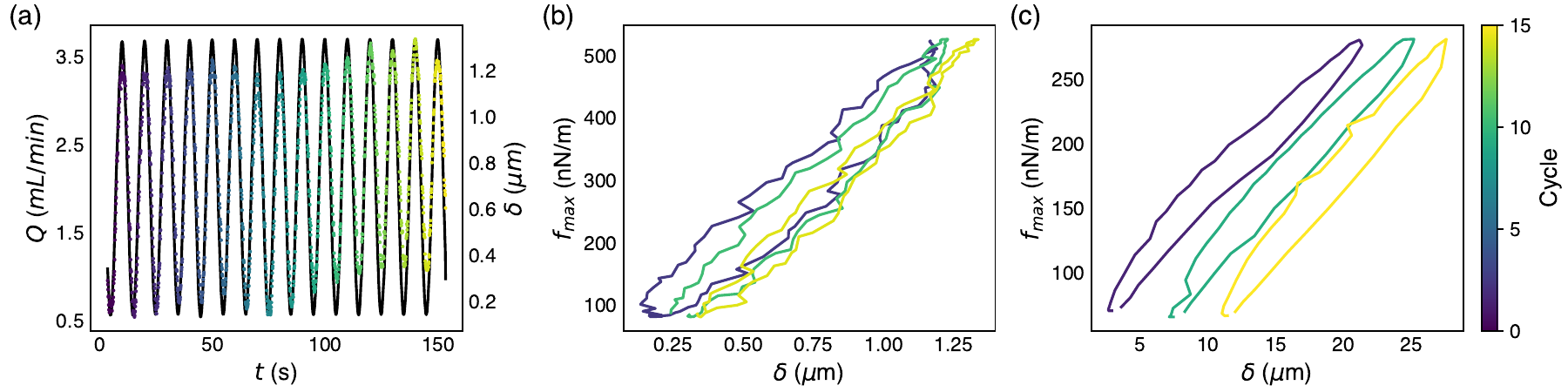}
    \caption{Cyclic loading reveals fiber properties. (a) An example of a sinusoidal cyclic loading test in a Ca-alginate fiber with 1\% NaCl. Flow rate is plotted in black, deflection is plotted as a dotted line in black. (b) Comparison of the loading/unloading from the first and last cycles. The deflection increases slightly over the course of the experiment. (c) A similar test on a Mg-alginate Fiber shows a much larger creep in deflection.}
    \label{fig:Sine Cycle Loading}
\end{figure}

An alternative way of exploring viscous behavior in a sample is using cyclic loading. On the macroscale, a technique called dynamic mechanical analysis (DMA) applies an oscillatory force on a material and measures the resulting deformation\cite{menard_dynamic_2020}. For viscoelastic samples, the deformation response is out of phase with the applied force, and the phase difference can be decomposed into an elastic and viscous component similarly to SAOS measurements performed on a rheometer.

Figure \ref{fig:Sine Cycle Loading} demonstrates DMA tests at a frequency of 0.1 Hz performed on two different alginate fibers. Figure \ref{fig:Sine Cycle Loading}a shows the raw data of one such experiment on a Ca-alginate fiber. The color gradient line shows the flow rate oscillating between 600 and 3600 $\mu$L/min 15 times over the course of 150 s, where the color gradient shows the time or cycle number. The dotted black line shows fiber deflection, which tracks closely with flow rate.
The sine curves in Figure \ref{fig:Sine Cycle Loading}a, representing the applied stress and the material response, are offset by a phase angle $\phi$.  However, $\phi$ is small, and not very apparent by visual inspection of Figure \ref{fig:Sine Cycle Loading}a.  Nonetheless we can fit both stimulus and response to sine curves in Figure \ref{fig:Sine Cycle Loading}a and obtain $\phi\sim11^{\circ}$.

The phase offset between the flow rate and the deflection becomes more apparent in the Lissajous curves seen in Figure \ref{fig:Sine Cycle Loading}b. In large amplitude oscillatory shear (LAOS) measurements, Lissajous curves plot the oscillatory stress as a function of the oscillatory strain or strain rate.  In the FIBR context, the oscillation of the applied hydrodynamic force causes an oscillation of the deflection, and plots of $\delta(f_{max})$ appear as closed loops.
The plot shows the force and deflection measurements from three sinusoidal cycles at the beginning, middle, and end of the experiment in purple, green, and yellow colors respectively. Each of the curves shows the oval shape characteristic of a viscoelastic material. Over the course of the experiment, the fiber shows slight net movement into the capillary of $\sim$0.2 $\mu$m. Figure \ref{fig:Sine Cycle Loading}c shows the Lissajous curves of a similar cyclic loading experiment performed on a softer Mg-alginate fiber. This fiber demonstrates the same characteristic oval shape.  
However, there appears to be a net translation in this sample, where the fiber moves nearly 10 $\mu$m into the capillary over the course of the experiment. This drift may reflect a combination of viscoelastic deformation and slow settling or slip at the capillary rim over the course of the 15 measurement cycles. Regardless of this observed translation, the slope of $f_{max}(\delta)$ remains constant, especially in later cycles.

The curves in Figure \ref{fig:Sine Cycle Loading}b are less smooth than those in Figure \ref{fig:Sine Cycle Loading}c. This is largely due to the challenge of precisely determining the rod position below the light diffraction limit for the fiber in Figure \ref{fig:Sine Cycle Loading}b that bends $<$1 $\mu$m, while the fiber in Figure \ref{fig:Sine Cycle Loading}c bends 20 $\mu$m.

Lissajous plots provide the phase offset and the relative contribution of the storage and loss moduli to the signal.  In LAOS, the volumetric energy stored during the measurement is proportional to an area of the curve: the change in applied stress multiplied by the change in strain.  The area within the curves is proportional to the volumetric energy lost.  The ratio of the loss modulus $G''$ to the storage modulus $G'$ is $\tan{\phi}$. The quantities obtained in FIBR are $f_{max}$ and $\delta$.  Converting to stress and strain requires dividing $f_{max}$ and $\delta$ by $R$ and $d$, respectively.  The dimensions of the area under the curve of $f_{max}(\delta)$ are of force rather than modulus.  However, the ratio of the two areas still provides $\tan{\phi}$. For the Ca-alginate and Mg-alginate samples shown in Figures \ref{fig:Sine Cycle Loading}b and \ref{fig:Sine Cycle Loading}c, $\phi=11.8^{\circ}$ and $\phi=10.3^{\circ}$, respectively.  

We can compare the results with SAOS performed at a fixed strain over a range of frequencies. Unlike SAOS, FIBR DMA is performed at a controlled frequency but without a fixed strain: measurement of the strain, or deflection, provides the material properties.  Regardless of this difference in the manner of measurement, we obtain remarkably similar results for the Ca-alginate fiber: $\phi=11.8^{\circ}$ from the FIBR DMA at $\omega=0.1$ Hz, and $\phi=11.4^{\circ}$ from SAOS at $\omega=0.1$ rad/s. In the softer Mg-alginate fiber sample, the SAOS measurement is $\sim30$\% less than the FIBR DMA result: $\phi=10.3^{\circ}$ by FIBR compared to $\phi\sim7^{\circ}$ from SAOS.  This discrepancy may be due to the fact that strain cannot be controlled in the FIBR method.  At the same time, sample-to-sample variations in $\phi$, even in well defined grades of alginate in solution, can vary by $\sim20$\% \cite{fu2010rheological}.

A slight sigmoidal shape in $f_{max}(\delta)$, or deviation from an ellipse, appears in Figure \ref{fig:Sine Cycle Loading}c. This might indicate a degree of strain stiffening. It may also be attributed to a deviation from a purely sinusoidal applied force.  At higher pressures, the vacuum takes time to achieve the set pressure, and pressure can lag behind the set pressure slightly, by as much as $\sim1$ s.  While this lag does not affect steady measurements, it may cause deviation from a perfect sine wave in pressure, and therefore $Q$, at the top of the cyclic loading curves like the one shown in Figure \ref{fig:Sine Cycle Loading}a.

\subsubsection{Failure}

\begin{figure}[htbp]
    \centering
    \includegraphics[width=1\linewidth]{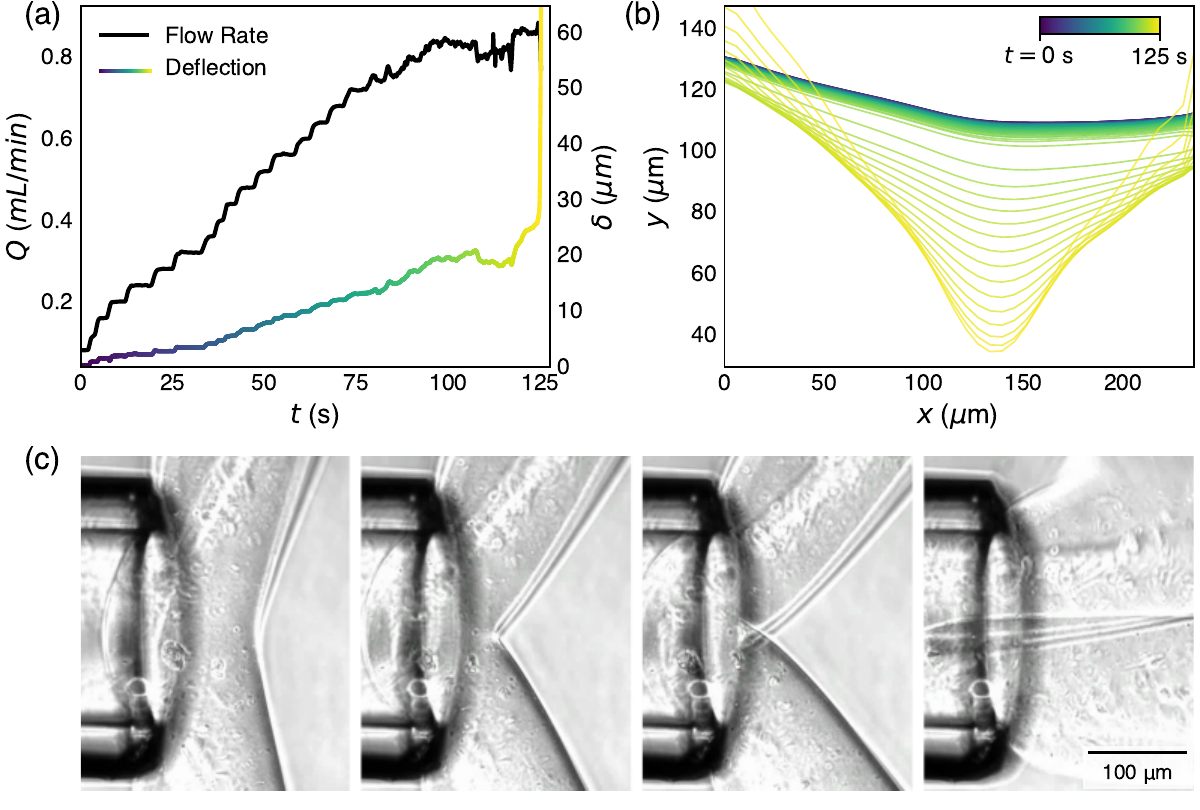}
    \caption{At a critical flow rate, the rod continues to deflect until it is pulled into the capillary, resulting in fiber failure. (a) Flow rate and deflection of a Mg-alginate fiber on a 200 $\mu$m capillary. Deflection increases rapidly at $\sim$120 s once flow rate is set to $\sim$850 $\mu$L/min. (b) Fitted contours show the shape change that rapidly happens at the end of the experiment. Color indicates the time: blue is $t=$ 0 s, yellow is $t=$ 125 s. (c) Microscopy showing the real-time rapid shape change of the Mg-alginate fiber during failure (Multimedia available online).}
    \label{fig:failure}
\end{figure}

In addition to time-dependent viscous measurements, the capillary bending technique can also measure other material characteristics, such as failure points. Figure \ref{fig:failure} shows one example of buckling failure of a Mg-alginate fiber  on a 200 $\mu$m capillary.
Figure \ref{fig:failure}a shows the raw data from the experiment. Flow rate increases from 100 to 800 $\mu$L/min over the course of 100 s. This results in elastic-like deflection in the fiber up to 20 $\mu$m. With this capillary, $>$800 $\mu$L/min is near the maximum possible flow rate and the flow rate is not as steady as we approach the limits of the vacuum pump. Nevertheless, as soon as the flow rate reaches 850 $\mu$L/min, the fiber moves dramatically. Within a couple of seconds, the fiber deflection more than triples and a significant crease develops in the microscopy. Finally, the fiber folds in half and is pulled into the capillary. Given the measured flow rate at the onset of failure, we estimate an applied stress $\sim120$ Pa causes this sample to bend catastrophically.  Notably for this sample, the fiber diameter is approximately the same size as the capillary so that significant squeezing must occur for the capillary to enter the capillary once folded in half. The video microscopy of this collapse, starting at $t=124$ s and continuing to $t=127$ s is seen in Figure \ref{fig:failure}c (Multimedia available online). Because in this case the fiber blocks a substantial fraction of the capillary entrance, the sample perturbs the entrance flow more strongly than assumed in the analytical model. In addition, the deformation leading to failure lies beyond the small-deflection regime. Therefore, we do not use this experiment to extract a modulus; rather, the imposed flow rate provides only an estimate of the nominal hydrodynamic load at which failure occurs.

The range for measuring failure stress with the FIBR rheometer is significantly narrower than the range for measuring the elastic modulus, due to the much larger bending that occurs during failure. While we can determine an elastic modulus with fiber movement of $<$5 $\mu$m, fiber failure means significant deflection. Nevertheless, we still see failure modes in a range of materials. In addition to the Mg-alginate fiber in Figure \ref{fig:failure}, we have observed similar buckling failure in our alginate rods (cross-section $40\times25$ $\mu$m) measured on a 600 $\mu$m capillary.  Young et al. report similar failure in some diatom samples, though do not discuss using this failure point as a means of characterizing samples \cite{young_quantifying_2012}.

\subsection{Experimental Limits and Challenges}

\begin{table}[htbp]
    \centering
    \setlength{\tabcolsep}{12pt}
    \renewcommand{\arraystretch}{1.3}
    \begin{tabular}{lll}
        \toprule[1.5pt]
        Parameter & Exp Min & Exp Max \\
        \midrule[1pt]
        $\delta$ & 1 $\mu$m &  100 $\mu$m \\
        $d$ &  5 $\mu$m &  330 $\mu$m \\
        $L$ &  200 $\mu$m &  2000 $\mu$m \\
        $Q$ & 50 $\mu$L/min & 15000 $\mu$L/min \\
        $\eta$ & 0.9 mPa s & 0.9 mPa s \\
        \bottomrule[1.5pt]
    \end{tabular}
    \caption{The range of experimental parameters of the capillary bending technique as performed in this paper. The actual experimental limits may vary by up to an order of magnitude beyond what is done here.}
    \label{tab:ExpLimits}
\end{table}

Table \ref{tab:ExpLimits} shows the range of parameters used in this study. Measurable deflection $\delta$ and fiber diameter $d$ are limited by the diffraction limit in light microscopy. The upper limits for these will depend on the capillary size and the small-deflection assumption in the Euler beam theory. Glass capillaries can be bought at a variety of sizes and glass pulling can produce capillaries with internal diameters down to single micrometers. Additionally, the technique, in theory, should work for arbitrarily large sizes, although the pump and imaging requirements for diameters above several millimeters may require different equipment.  For larger samples, which might not be suitable for microscopy, the use of two or multiple camera views might be useful to monitor the position of the sample through the course of an experiment. The achievable flow rates will vary based on the size of the capillary and the capacity of the pump. Our Fluigent vacuum pump can achieve $\sim-900$ mbar of pressure. For the 600 $\mu$m capillary, this allows for flow rates up to $\sim$1 mL/min, while the 2000 $\mu$m capillary can draw $\sim$15 mL/min at the same pressure. The lower flow rate limit depends on the resolution of the pump and on the ability to hold the fiber in place. We use a low flow rate to hold the fiber in position against the capillary (50-500 $\mu$L/min depending on the fiber size), although alternate methods for holding the fiber in position may allow for lower flow rates. Lastly, all experiments in this paper use water at room temperature, which has a viscosity of $\sim$0.9 mPa s. The experimental technique is not limited to this however, and adding solutes to the water, changing the solvent, and changing the temperature can all change the viscosity substantially. The corresponding values of \(\mathrm{Re}_f\) are listed in Table~S1.

With the experimental ranges given in Table \ref{tab:ExpLimits}, we can measure flexural stiffnesses in the range $10^{-18}< EI < 10^{-12}$ Pa m$^4$. With differently sized capillaries and by varying other parameters as discussed above, one should be able to extend that range by at least an order of magnitude on either end. In Table \ref{tab:Bending Results}, we show the resulting elastic moduli ranging from $10^2$ to $10^8$ Pa. With sufficiently small fibers it may be possible to measure moduli several orders of magnitude higher. Consider the polyester fiber, which have a diameter of 28 $\mu$m. A fiber with a similar effective stiffness but a diameter of 2.8 $\mu$m would have a modulus of $\sim$500 GPa. Softer fibers are also possible, although manipulating softer fibers may damage them. The $E=$ 300 Pa Mg-alginate fibers are so soft that manipulating them with tweezers often collapses or breaks the fibers.

Within the experimental limits, we found the most common causes of error to be improper fiber alignment. Fibers must be positioned across the center of the capillary to properly quantify the span and ensure an appropriate force distribution. The fiber also must touch the capillary on both sides of the span. We rely on a low flow rate to pull the fiber onto the capillary, both to secure contact on both sides and position it on the longest span of the capillary. We have also found that longer fibers are more difficult to position in this way, and therefore harder to measure. Short fibers, $<5\times$ the capillary diameter, are able to be pulled into position more easily through the force of the flow alone. Longer fibers may require manual manipulation for positioning. Additionally, small manipulations at the loose end of a fiber can produce exaggerated movements down the length of the fiber. This is particularly a concern at high flow rates, as there may be substantial currents produced within the well by the water resupply line, which can dislodge the fiber from the capillary. Thus, choosing an appropriate capillary size and, if possible, reducing the length of the fiber will lead to easier and more accurate measurements.

We also note here some additional limitations of the presented theoretical description and of the experimental implementation. The current mathematical model relies on several idealizing assumptions, including the small-deflection limit, negligible sample porosity, a laterally centered fiber, and simply supported, no-slip boundary conditions at the capillary edge. In addition, the hydrodynamic description is derived in the Stokes-flow limit, i.e. for negligible fluid inertia. In practice, however, the experimentally observed force--deflection data remain close to linear over a broader range of flow rates than would be expected from a strictly asymptotic $Re\ll 1$ argument alone, as seen in Fig.~\ref{fig:elastic}b, suggesting that the method remains empirically robust into a finite-Reynolds-number regime. Nevertheless, under such conditions the present model should be regarded as an effective description, and more refined hydrodynamic corrections may be required for precise quantitative treatment at higher flow rates.

Moreover, the beam description is formally derived in the small-deflection limit, whereas the largest measurements in this study reach $\delta/(2R)\approx 0.15$.  The linearity of Fig.~\ref{fig:elastic}b suggests that the model remains useful as an effective description over this range, but it should not be interpreted as a fully nonlinear beam treatment. At larger deformations several nonlinear effects may enter, including geometric nonlinearity, changes in the hydrodynamic load distribution as the fiber shape evolves, and potentially nonlinear viscoelastic response of the material itself \cite{kamrin2012soft}.  For this reason, we do not interpret the large-amplitude failure observation with the same small-deflection elastic model.

We expect the method to be most accurate for fibers that are approximately uniform along the measured span. Surface roughness, local defects, weak points, or strong variations in cross-section can alter the local drag or bending resistance and may localize curvature, such that the inferred modulus reflects an effective bending response of the tested segment rather than an intrinsic material constant of an ideal homogeneous beam. At larger deformations, nonlinear effects and deviations from the idealized beam model may also become apparent. For instance, with significant deflections or under cyclic loading, contact friction at the capillary edge or local fiber slip might no longer be negligible, thereby modifying the effective boundary conditions and the total length of the fiber.

Furthermore, soft hydrogels are often porous; potential fluid flow through the material structure would decrease the hydrodynamic drag force compared to a non-porous cylinder. This implies that the current theory might slightly over-predict the bending modulus for highly permeable materials. A practical criterion for neglecting porosity is that the Darcy flux through the fiber should remain small compared with the externally imposed flux through the capillary. For the microfluidic alginate rods, we observe complete channel occlusion together with no detectable leakage above approximately $1~\mu$L/min at pressure differences of order 2~bar.  This implies an upper bound of permeability of $k \lesssim 8\times 10^{-14}\,\mathrm{m}^2$, or about 80~mD, based on Darcy's equation. This suggests that the flow through the sample is estimated to be small, 1\% or less, on the scale relevant to the present measurements.  However, materials with substantially higher permeability would require an explicit poroelastic correction. The cases of large deformation involving viscoelastic effects, probed in oscillatory loading, are therefore treated phenomenologically.

\section{Conclusion}

The FIBR rheometer is uniquely capable of measuring small, soft, hydrated fibers. The technique is able to measure smaller and softer materials than traditional DMA instruments while avoiding the expense and difficulty in data analysis of an AFM. Simple elastic measurements with the capillary bending technique can be done very inexpensively with common equipment. While we use specialty pressure-driven pumps designed for microfluidics, any syringe pump with withdrawal capabilities is sufficient.
For time-dependent measurements, changes in flow rate must be synchronized with the video recording; alternatively, tracer particles can be added to the water to determine flow rate via microscopy. More advanced dynamic mechanical analysis requires a pump configured to provide sinusoidal flow, and the range of accessible frequencies will depend on the pump's capabilities and flow rate resolution.

One limitation of the FIBR rheometer is that it controls for flow rate, and therefore stress, but not strain. Many applications, such as relaxation measurements, require achieving a specified strain. There may be an adaptation to the capillary bending rheometer which allows for controlling strain. With proper equipment and real-time image analysis, a feedback loop can adjust the stress in real time to achieve and maintain fiber deflection at a certain value. The delay inherent in any feedback loop likely limits this approach to equilibrium measurements and very low frequency oscillations.

The FIBR rheometer fills a critical gap for hydrated materials by operating in a submerged environment with non-destructive forces. This makes it particularly well-suited for biological materials and materials engineered for bio-environments, where maintaining hydration is essential for accurate characterization.
Biological materials and cells, such as neurons, may be particularly amenable to measurement with this technique. Even beyond dynamic mechanical measurements, the hydrodynamic model and experimental setup can probe the active response of biological materials, that is, how living cells respond to forces in their environment.  Many active and biological materials respond to external mechanical forcing by growing and/or adjusting their own material properties.  FIBR could be adapted to measure such emergent states as a function of externally applied forces.

\section*{Supplemental Material}

See the supplemental material for detailed hydrodynamic force calculations, comparison of exact and approximate analytical solutions, SAOS rheology data for bulk alginate gels, and alternative viscoelastic model fits for the creep data.

\section*{Acknowledgments}
BTS and SMH are supported by the NSF CBET CAREER 2239742 to SMH. The work of MC and ML was supported by the National Science Centre of Poland Sonata Bis grant no. 2023/50/E/ST3/00465 to ML.

We thank William Lee for his assistance in manufacturing and measuring alginate fibers.

\section*{Conflict of Interest}
The authors have no conflicts to declare.

\section*{Data Availability Statement}
The data that support the findings of this study are available from the corresponding author upon reasonable request.

\bibliography{references, theory_ref}

\clearpage
\setcounter{figure}{0}
\renewcommand{\thefigure}{S\arabic{figure}}
\setcounter{equation}{0}
\renewcommand{\theequation}{S\arabic{equation}}
\setcounter{table}{0}
\renewcommand{\thetable}{S\arabic{table}}

\section{Supplemental Information}

\subsection{Image Analysis Pipeline}

\begin{figure}[h!]
    \centering
    \includegraphics[width=1\linewidth]{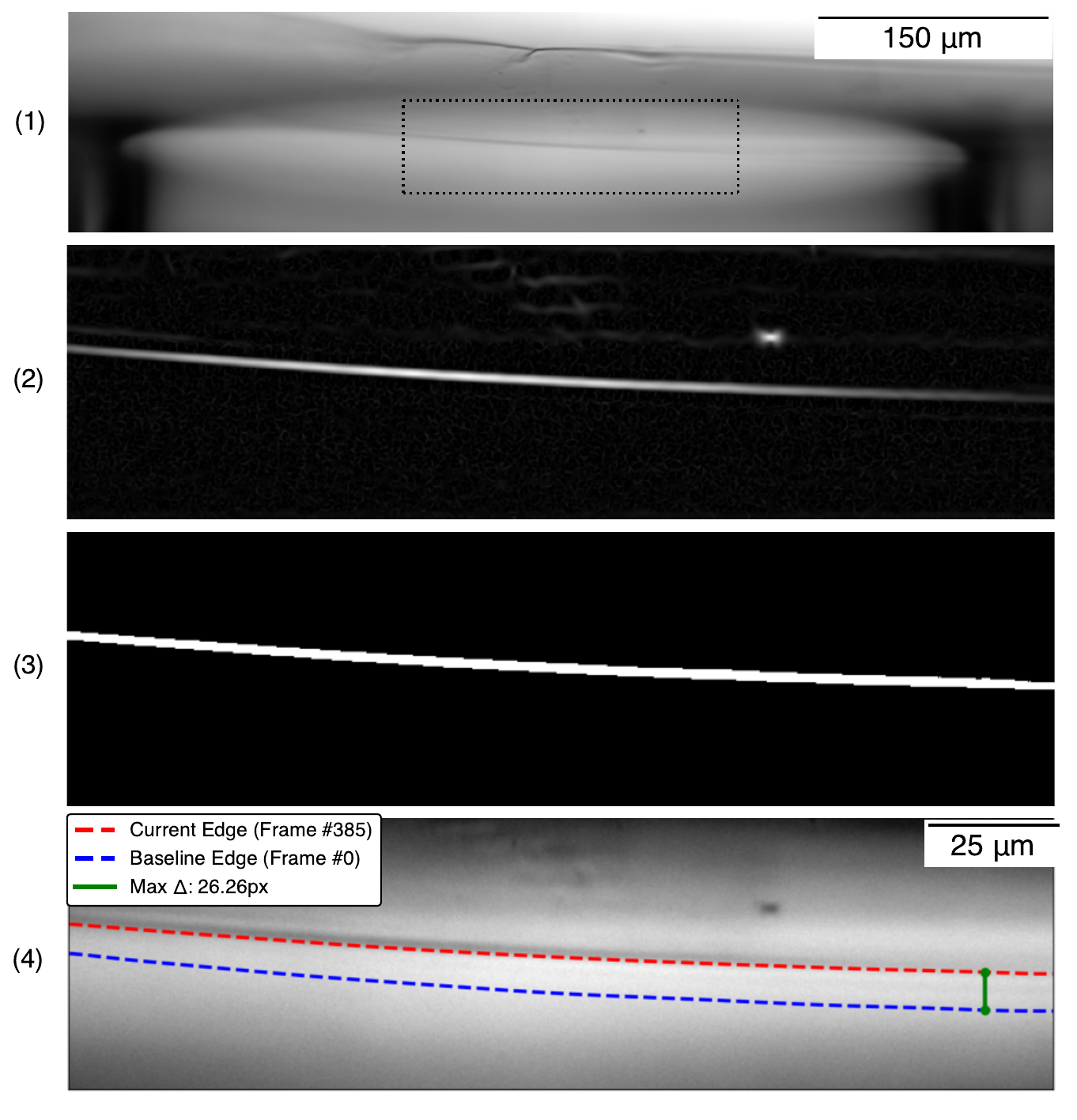}
    \caption{(1) shows the raw microscopy image corresponding to frame 385. (2)-(3) show the image analysis steps performed on the region within the dashed box in (1). (2) shows the results of applying the Sato on the raw image: bright regions appear where ridges or edges are found. (3) is a binarized version of (2).  (4) shows the two contours found in Frames 0 and 385 overlaid on top of the raw microscopy image.}
    \label{fig:SIVidAnalysis}
\end{figure}

Figure \ref{fig:SIVidAnalysis} shows the steps of the image analysis pipeline for non-fluorescent samples.  The sample shown is an Ca-alginate fiber of diameter 70$\mu$m measured on a capillary with inner diameter 0.6mm.  The entire diameter of the sample fiber is visible in the raw microscopy image in (1).  To find the inner edge of the sample, we focus on the region of interest (ROI) shown in the dashed box. Applying the Sato filter generates bright regions wherever ridges or edges are found, as seen in (2).  Quantitative contour finding requires a binarized image as input: (3) shows the binarized version of (2).  Contours are found in every frame of the microscopy video and compared to the location of the fiber edge in the initial frame 0 of the microscopy video to determine the deflection $\delta$. In (4), the contours found in frames 0, in blue, and 385, in red, are overlaid on top of the raw microscopy image.  The difference between the red and blue dashed contours is measured across the entire ROI. The vertical green line segment shows the location of the maximum difference $\delta = 8.4 $ $\mu$m.

\clearpage

\subsection{Establishing theoretical assumptions}

Table \ref{tab:ReTable} presents the maximal bending conditions observed from each material measured.  The parameters listed include the radius of the capillary used, $R$, the diameter of the sample material fiber, $d$, the maximum deflection measured $\delta_\mathrm{max}$.  These maximal deflections are always measured at the highest flow rates, $Q_\mathrm{max}$.  Thus, the conditions listed in Table \ref{tab:ReTable} represent those with the highest fiber Reynolds number \(\mathrm{Re}_f\).  This table provides the experimental parametric basis for the theoretical assumptions that bending is small and inertial effects are insignificant.  Most materials bend such that $\delta_\mathrm{max}/2R\lesssim 5$\%.  Only the alginate rods exhibit a greater deflection, $\delta_\mathrm{max}/2R\sim 11$\%. The maximum \(\mathrm{Re}_f=41\) is observed for the Ca-alginate fibers in NaCl.

\begin{table}[htbp]
    \centering
    \setlength{\tabcolsep}{12pt}
    \renewcommand{\arraystretch}{1.3}
    \begin{tabular}{lllllll}
        \toprule[1.5pt]
        Material & $2R$ & $Q_\mathrm{max}$ & $d$ & $\delta_\mathrm{max}$ & $\mathrm{Re}_{f}$ \\
         & ($\mu$m) & ($\mu$L/min) & ($\mu$m) & ($\mu$m) & \\
        \midrule[1pt]
        Downy feather barbule & 200 & 822 & 5.3 & 11.1 & 4 \\
        Polyester Fiber & 2000 & 15103 & 26.8 & 94.9 & 4 \\
        Ca-alginate (2\%:2\%) & 1000 & 3659 & 221.6 & 4.5 & 29 \\
        Ca-alginate (2\%:2\%) + 1\%NaCl & 2000 & 14071 & 326.9 & 16.7 & 41 \\
        Mg-alginate (2\%:2\%) & 600 & 313 & 151.3 & 13.4 & 5 \\
        Alginate Rods & 200 & 632 & 25.0 & 29.6 & 14 \\
        \bottomrule[1.5pt]
    \end{tabular}
    \caption{Experimental conditions leading to maximum bending in each material, using a capillary of radius $R$ to measure a fiber of diameter $d$. The maximum flow rate $Q_\mathrm{max}$ leads to the maximum bending deflection $\delta_\mathrm{max}$, and also provides the maximum $\mathrm{Re}_{f}$.}
    \label{tab:ReTable}
\end{table}

\clearpage

\subsection{Rotation of Free Ends}

\begin{figure}[h!]
    \centering
    \includegraphics[width=1\linewidth]{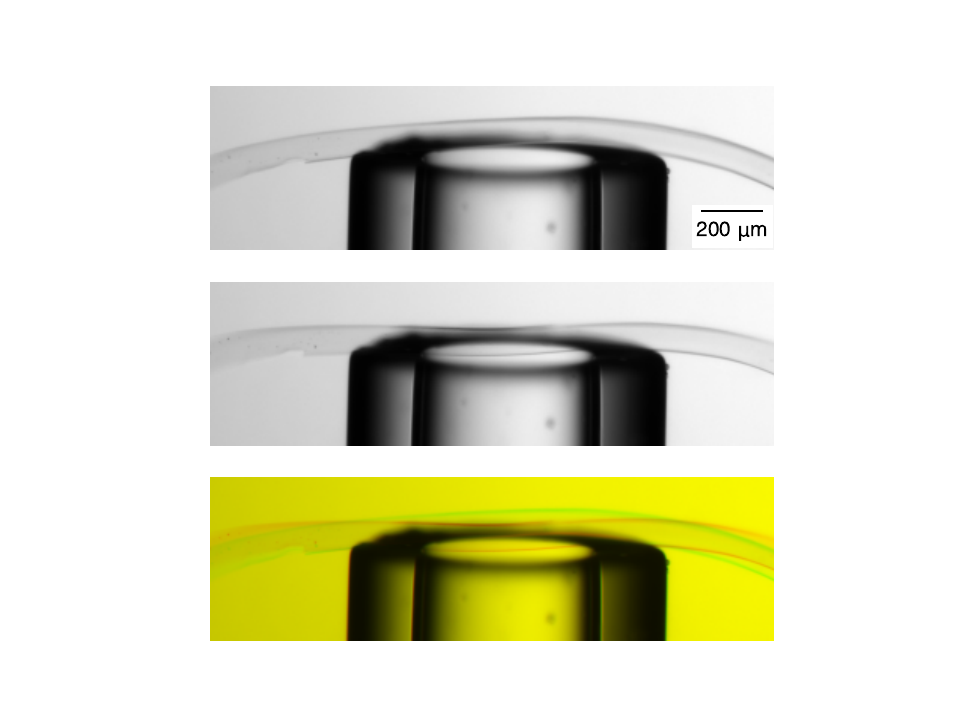}
    \caption{The top and center panels show raw microscopy images of a Ca-alginate fiber measured on a capillary with inner diameter 0.6mm at applied flow rates of 200 and 1600 $\mu$L/min.  The bottom panel shows a composite of the two images.}
    \label{fig:SIRotatingEnds}
\end{figure}

Figure \ref{fig:SIRotatingEnds} shows the position of the fiber both inside and outside the capillary during bending. The sample shown is a Ca-alginate fiber of diameter 100 $\mu$m measured on a capillary with inner diameter of 600 $\mu$m.  In the top image, the applied flow is 200 $\mu$L/min.  In the center image, the applied flow is 1600 $\mu$L/min.  The bottom image in Figure \ref{fig:SIRotatingEnds} represents a composite of the two images above.  The green contour is the initial shape from the top image, while the orange contour is the shape from the center image.  The fiber rotates about its resting positions on the capillary as it deflects, which justifies the boundary conditions used in the theoretical formulation.

\subsection{Calculations of the hydrodynamic load on the fibers}

For microscale objects, the flow $\bm{u}$ around the whole fluid domain is governed by the incompressible Stokes equations
\begin{eqnarray}
    \eta \nabla^2 {{\bm{u}}} -\nabla p + {{\textbf{f}}}&=&0,\\
    \nabla\cdot{{\bm{u}}}&=&0,
\end{eqnarray}
where $\eta$ is the dynamic viscosity and $p$ is the pressure field. If a force ${\bm{F}}$ acts on a particle immersed in Stokes flow, it will move with a velocity ${{\bm{u}}}$ according to the relationship
\begin{equation}
      {{\bm{u}}} = {\bm{{\mu}}}\cdot{\bm{F}},
      \label{eq:mobility}
\end{equation}
where ${\bm{{\mu}}}$ is the mobility matrix \cite{kim2013microhydrodynamics}. The linear relationship also holds for a multiparticle system, where for $N$ particles the vectors $ {{\mathbf{u}}}$ and $ {{\bm{F}}}$ take the form of $ {{\bm{u}}} = [\bm{u}_1, \ldots,\allowbreak \bm{u}_N]$ and $ {{\bm{F}}} = [\bm{F}_1, \ldots,\allowbreak \bm{F}_N]$, and ${\bm{{\mu}}}$ is a $3N\times 3N$ matrix composed of $3\times 3$ blocks $\bm{\mu}_{ij}$. A sub-matrix $\bm{\mu}_{ij}$ relates the velocity of particle $i$ to the forces acting on particle $j$. In Stokesian Dynamics, and in the absence of ambient flow, the configuration of the system ${{\bm{R}}} = [\bm{R}_1, \ldots,\allowbreak \bm{R}_N]$ evolves according to
\begin{equation}
    \td{\bm{R}}{t} = \bm{\mu}\cdot \textbf{F}.
    \label{eq:mobilitymatrix}
\end{equation}
In general, the mobility matrix $\bm{\mu}$ depends on the positions of all particles. Assuming that the particles interacting in the fluid are identical spheres of radius $a$, we shall use the pairwise Rotne-Prager-Yamakawa approximation to resolve the hydrodynamic interactions. Within this model, we assume that $\bm{\mu}_{ij}$ depends only on the positions of the particles $i$ and $j$ via
\begin{equation}
  \frac{\bm{\mu}_{ij}}{\mu_0} =
    \begin{cases}
      \displaystyle  \frac{3a}{4{R_{ij}}}\left[\left(1+\frac{a^2}{3R_{ij}^2}\right)\mathds{1} + \left(1- \frac{2 a^2}{R_{ij}^2}\right)\hat{\bm{R}}_{ij}\hat{\bm{R}}_{ij} \right] & \text{for $R_{ij} \geq 2a$,} \\[3ex]
      \displaystyle \left(1-\frac{9R_{ij}}{32a}\right) \mathds{1} +\frac{3 R_{ij}}{32a }\hat{\bm{R}}_{ij}\hat{\bm{R}}_{ij} & \text{for $ 2a \geq R_{ij} >0$},\\
    \end{cases}
\end{equation}
where $a_i$ denotes the radius of the $i$-th bead, $R_{ij}=|\bm{R}_{ij}|$ is the distance between the $i$-th and $j$-th bead, and $\hat{\bm{R}}_{ij}$ is the unit vector from bead $i$ to bead $j$. Here,  $\mu_0=(6\pi\eta a)^{-1}$ is the Stokesian mobility of an isolated spherical particle. Within this approximation, the velocity of the sphere $i$, $\bm{U}_i = \text{d}\bm{R}_i/\text{d}t$, is given by
\begin{equation} \label{eq:hydromotion}
    \bm{U}_i = \sum_{j} \bm{\mu}_{ij}\cdot \bm{F}_j = \mu_0 \bm{F}_i + \sum_{j\neq i} \bm{\mu}_{ij}\cdot \bm{F}_j,
\end{equation}
where we singled out the self-term $i=j$ being the Stokes velocity.

We now introduce the bead-model, in which a fiber is represented by a collection of $2N+1$ identical, possibly overlapping spheres, located at the points $\bm{R}_i$, with $i=1,\ldots,2N+1$, initially aligned together in the form of a straight rod \cite{manghi2006hydrodynamic,wang2020hencky}. The diameter of each bead is $2a$, and the centers of the subsequent beads are separated by a distance $l_0$.

The experiments span a range of aspect ratios. We distinguish two regimes by comparing the diameter of the fiber, $d=2a$, to the diameter of the capillary. This aspect ratio $R/d$, determines whether the fiber is `thick' or `thin'. We model these limits differently: for thick fibers we assume that the capillary diameter is filled with overlapping beads, while for thin fibers we represent the fiber by a string of non-overlapping and touching beads.

The idea behind the calculation is the following. Because in the initial state the flow is perpendicular to the vector ${\hat{\bm{R}}}_{ij}$, for every $ij$ pair the dyadic part of the mobility matrix does not contribute to the final expression. For the perpendicular direction, along the axis of the capillary $z$, we can write down
\begin{equation}
    U_i^z = \mu_{ij}^{zz}F_j^z
    \label{eq:rpy_z}
\end{equation}
Initially, thin filaments are considered. The thin filament is modeled as a collection of $2N+1$ spherical beads of radius $a$, such that the centers of the outermost beads are located at positions $x = \pm R$. The beads are arranged so that each one touches only its immediate neighbors, and the contacts occur at a single point. In such geometry, the $zz$ components of the mobility matrix take the form of
\begin{equation}
  \mathbf{\mu}_{ij}^{zz} =
    \begin{cases}

      \displaystyle  \frac{1}{16\pi\eta{a}\lvert i-j\rvert}\left(1+ \frac{1}{6\lvert i-j\rvert^2} \right) & \text{for $i\neq j$,}\\[3ex]

      \displaystyle \frac{1}{6\pi\eta{a}} & \text{for $ i = j$.}

      \label{eq:RPY_ij_thin}
    \end{cases}
\end{equation}
The main focus is the velocity felt by the middle bead  ($i = N+1$) and the total force acting on the middle bead. Using equations  \eqref{eq:RPY_ij_thin} and \eqref{eq:force_quartic}, we find
\begin{eqnarray}
         &&U_\text{max} = \sum_{i=1}^{2N+1}\mathbf{\mu}_{Nj}^{zz}F_j = \nonumber\\
         &&
\frac{F_\text{max}}{16\pi\eta a}\sum_{i\neq N}^{2N+1}\left(\frac{1}{\lvert i-N\rvert} + \frac{1}{6\lvert i-N\rvert^3} - \frac{\lvert i-N\rvert^3}{N^4}- \frac{\lvert i-N\rvert}{6N^4} \right) +  \\
         &&+\frac{F_\text{max}}{6\pi\eta a} = \frac{F_\text{max}}{8\pi\eta a} \sum_{k=1}^{N}\left(\frac{1}{k}+\frac{1}{6k^3} - \frac{k^3}{N^4} -  \frac{k}{6N^4} \right) + \frac{F_\text{max}}{6\pi\eta a}.
         \nonumber
         \label{eq:rpy_sum2}
\end{eqnarray}

The exact analytical result of the above sum is given by:

\begin{equation}
    F_\text{max} = \frac{8\pi\eta a U_\text{max}}{\displaystyle H_N + \frac{1}{6}H^{(3)}_N+\frac{13}{12}-\frac{1}{2N}-\frac{1}{3N^2}-\frac{1}{12N^3}}.
    \label{eq:force_exact}
\end{equation}
where $H_n$ and $H^{(3)}_n$ denote the harmonic numbers and the harmonic numbers of the second order, respectively.

\begin{figure}[h]
    \centering
    \includegraphics[width=0.6\linewidth]{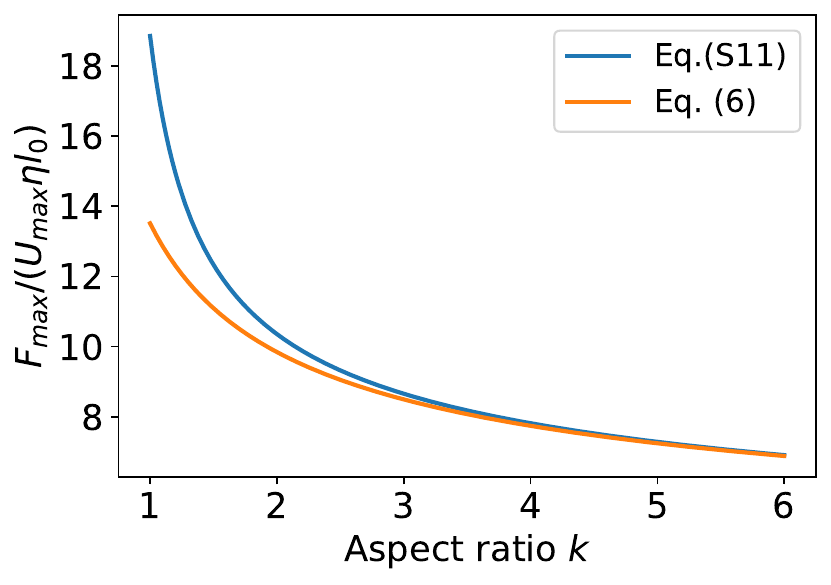}
    \caption{Comparison of the dimensionless force values obtained from the exact (Eq.~\eqref{eq:force_exact}) and approximate (Eq.~\eqref{eq:force_thin}) solutions. We observe that the differences are generally small and decrease as the aspect ratio $k$ increases.}   \label{fig:rpy_exact_vs_simplified}
\end{figure}

For sufficiently large $N$, and therefore large aspect ratio $k$, we can simplify Eq. \eqref{eq:force_exact} and divide by the characteristic length $l_0 = 2a$ to obtain equation \eqref{eq:force_thin}.
The approximate solution in equation \eqref{eq:force_thin} provides $f_{max}$, a force per unit length.  To compare the exact analytical result in Eq. \eqref{eq:force_exact} to Eq. \eqref{eq:force_thin}, we also divide $F_{max}$ by $l_0$
\begin{equation}
    f_\text{max} = \frac{4\pi\eta U_\text{max}}{\displaystyle H_N + \frac{1}{6}H^{(3)}_N+\frac{13}{12}-\frac{1}{2N}-\frac{1}{3N^2}-\frac{1}{12N^3}}.
    \label{eq:forceperlength_exact}
\end{equation}
The differences between the forces calculated using the exact and approximate solutions are small, as shown in Figure \ref{fig:rpy_exact_vs_simplified}. Therefore, the approximate version is used to interpret experimental measurements.

In the case of thick filaments, we proceed differently. Let us assume that the ratio of the tube radius $R$ to the sphere radius $a$ is equal to $k$, where $k \in [1,2]$. We then construct a filament from $2N+1$ overlapping beads such that the centers of the outermost beads are located at positions $x = \pm R$, and the distance between two neighboring beads is $R_{i,i+1} = ka/N$. The $zz$ components of the mobility matrix take the form \cite{zuk2014rotne}

\begin{equation}
  \mathbf{\mu}_{ij}^{zz} =
    \begin{cases}

      \displaystyle  \frac{1}{6\pi\eta{a}}\left(1- \frac{9k\lvert i-j\rvert}{32N} \right) & \text{for $i\neq j$,}\\[3ex]

      \displaystyle \frac{1}{6\pi\eta{a}} & \text{for $ i = j$.}

      \label{eq:RPY_ij_thick}
    \end{cases}
\end{equation}
Instead of force acting on the middle bead, we will focus on the force acting on the part of the middle bead of length $ka/N$:
\begin{eqnarray}
         &&U_\text{max} = \sum_{i=1}^{2N+1}\mathbf{\mu}_{Nj}^{zz}F_j = \nonumber  \\
         && \frac{f_\text{max}}{6\pi\eta a}\sum_{i\neq N}^{2N+1}\left(1 - \frac{9\lvert i-N\rvert}{32N} - \frac{\lvert i-N\rvert^4}{N^4} + \frac{9\lvert i-N\rvert^5}{32N^5} \right) + \frac{1}{6\pi\eta a} = \\
         &&\frac{f_\text{max}}{3\pi\eta a} \sum_{l=1}^{N}\left(1-\frac{9kl}{32N} - \frac{l^4}{N^4} +  \frac{9kl^5}{32N^5} \right) + \frac{F_0}{6\pi\eta a}.\nonumber
         \label{eq:rpy_sum}
\end{eqnarray}
Because the parameter $N$ is arbitrary here, we will assume that $N\ll 1$, and therefore we can write down approximated relationship given by Eq.~\eqref{eq:force_thick}.\\

Force profiles found in both regimes were placed in Eq.~\eqref{eq:euler_bernouli_beam} and integrated considering boundary conditions given by Eq.~\eqref{eq:euler_bernouli_bc}. The integration leads to the relationship in Eq.    \eqref{eq:euler_bernoulli_EI}.
By substituting the appropriate linear force densities and the corresponding second moments of area, we obtain expressions for $E$ in both the thin and thick regimes, and for both cylindrical and square cross-sections.  These expressions are given in equations \eqref{eq:E_thin_cylinder}-\eqref{eq:E_thick_square}. Substituting the exact expression for the linear force density for thin fibers from Eq.
~\eqref{eq:force_exact}
yields the following relation:
\begin{equation}
    E = \frac{\eta U_{max}}{\delta}\frac{256}{5}\frac{k^4}{\displaystyle H_k + \frac{1}{6}H^{(3)}_k+\frac{13}{12}-\frac{1}{2k}-\frac{1}{3k^2}-\frac{1}{1k^3}}.
    \label{eq:E_exact}
\end{equation}

For completeness, we have included all expressions for Young's modulus as a function of thickness and cross-sectional area in Table \ref{tab:E_equations}.

\begin{table}[ht]
\centering
\renewcommand{\arraystretch}{2.2}
\begin{tabular}{|l|c|c|}
\hline
\textbf{Cross-section shape} & \textbf{Thin fibres} & \textbf{Thick fibres} \\
\hline

\textbf{Circular} &
$\displaystyle
\frac{256}{5}\frac{U_{\text{max}}\eta}{\delta}\frac{k^4}{\log k + 1.86}
$ &
$\displaystyle
24\frac{\eta U_{\text{max}}}{\delta}\frac{k^3}{1-\frac{15}{64}k}
$ \\
\hline

\textbf{Square} &
$\displaystyle
\frac{48\pi}{5}\frac{U_{\text{max}}\eta}{\delta}\frac{k^4}{\log k + 1.86}
$ &
$\displaystyle
\frac{9\pi}{2}\frac{\eta U_{\text{max}}}{\delta}\frac{k^3}{1-\frac{15}{64}k}
$ \\
\hline

\textbf{Rectangular} &
$\displaystyle
\frac{48\pi}{5}\frac{U_{\text{max}}\eta}{\delta}\frac{R^4}{x^3 y}
\frac{1}{\log \!\left(\frac{R}{x^{1/2}y^{1/2}}\right)+1.86}
$ &
$\displaystyle
\frac{9\pi}{2}\frac{\eta U_{\text{max}}}{\delta}
\frac{R^3}{x^{5/2}y^{1/2}\left(1-\frac{15}{64}\frac{R}{x^{1/2}y^{1/2}}\right)}
$ \\
\hline

\end{tabular}
\caption{Expressions for $E$ for different fiber cross-sections and thickness regimes.}
\label{tab:E_equations}
\end{table}

\subsection{Bulk Rheology}

\begin{figure}[h!]
    \centering
    \includegraphics[width=1\linewidth]{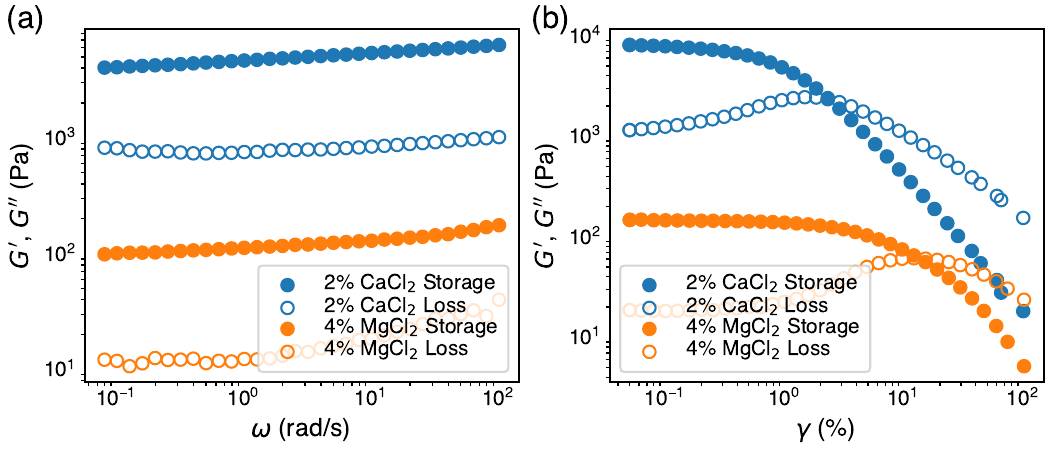}
    \caption{(Small) amplitude oscillatory shear rheology on bulk alginate gels.}
    \label{fig:SIRheology}
\end{figure}

Figure \ref{fig:SIRheology} shows the small amplitude oscillatory shear rheology of two alginate gels prepared in bulk, after being soaked in calcium and magnesium salt baths. Figure \ref{fig:SIRheology}a shows a frequency sweep at a constant strain of 0.1\%. Figure \ref{fig:SIRheology}b shows an amplitude sweep at a frequency of 5 rad/s. The average moduli across the entire frequency sweep of Figure \ref{fig:SIRheology}a are reported in the SAOS measurements in Table \ref{tab:Bending Results}.

Figure \ref{fig:SIRheology}a shows that $G'$ and $G''$ vary slightly when $\gamma=0.1$ \% and $\omega$ increases from 0.09 to 100 rad/s. As a result, $\phi$ decreases from $11.5^{\circ}$ to $9^{\circ}$ for the Ca-alginate fiber and increases from $\sim6.5^{\circ}$ to $13^{\circ}$ for the Mg-alginate fiber as $\omega$ increases.

\subsection{Alternate Creep Models}

\begin{figure}[h!]
    \centering
    \includegraphics[width=1\linewidth]{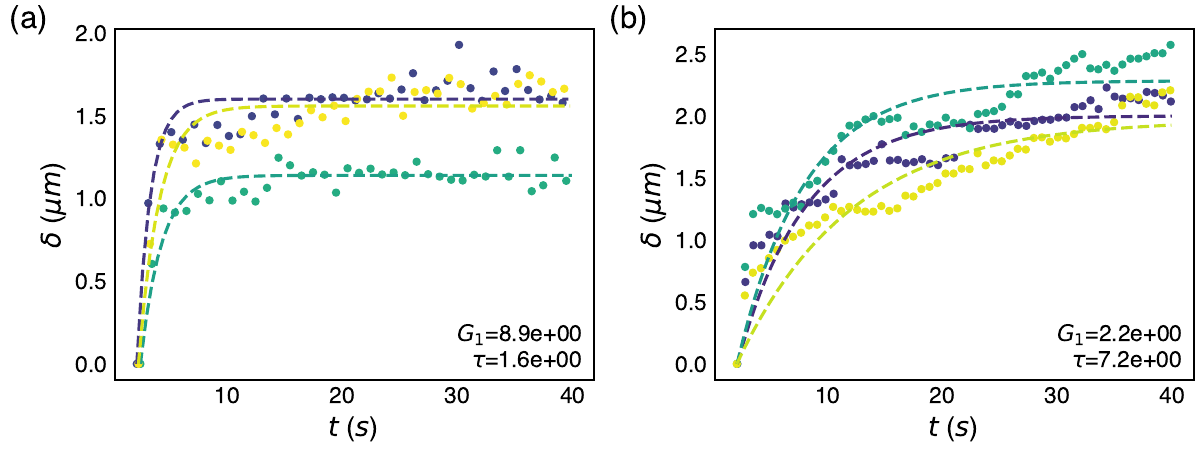}
    \caption{Kelvin-Voigt Fit to creep data.}
    \label{fig:SI_KV_Creep}
\end{figure}

Figures \ref{fig:SI_KV_Creep}a and \ref{fig:SI_KV_Creep}b show the creep data from Figures \ref{fig:Square Cycle Creep}a for a Ca-alginate sample, and \ref{fig:Square Cycle Creep}c for an Mg-alginate sample, respectively, as well as the fit of the Kelvin-Voigt model of viscoelasticity\cite{malkin_rheology_2022, li_facile_2017}.
The Kelvin-Voigt model behaves according to the following equation:
$$ J(t) = \frac{1}{G_1}\left(1-e^{-t/\tau}\right)$$
where $J$ is the viscoelastic compliance $\gamma(t)/\sigma_0$, $G_1$ is the shear modulus of the spring element in the model and $\tau$ is the decay time.

\end{document}